\documentstyle[12pt,aaspp4,epsbox]{article}

\lefthead{Kenji Bekki and Masashi Chiba}
\righthead{The formation of the Galactic stellar halo}

\begin{document}
\title{Formation of the Galactic Stellar Halo. I. 
Structure and Kinematics}

\author{Kenji Bekki} 
\affil{
School of Physics, University of New South Wales, Sydney 2052, Australia} 

\and

\author{Masashi Chiba}
\affil{
National Astronomical Observatory, Mitaka, Tokyo, 181-8588, Japan}

\begin{abstract}

We perform numerical simulations for the formation of the Galactic stellar
halo, based on the currently favored cold dark matter (CDM) theory of galaxy
formation. Our numerical models, taking into account both dynamical and
chemical evolution processes in a consistent manner, are aimed at explaining
observed structure and kinematics of the stellar halo in the context of
hierarchical galaxy formation. The main results of the present simulations
are summarized as follows.

(1) Basic physical processes involved in the formation of the stellar
halo, composed of metal-deficient stars with [Fe/H] $\le$ $-1.0$,
are described by both dissipative and dissipationless merging of subgalactic
clumps and their resultant tidal disruption in the course
of gravitational contraction of the Galaxy at high redshift ($z$ $>$ 1).

(2) The simulated halo has the density profile similar to the observed
power-law form of $\rho (r)$ $\sim$ $r^{-3.5}$, and has also the similar
metallicity distribution to the observations. The halo virtually shows
no radial gradient for stellar ages and only small gradient for metallicities.

(3) The dual nature of the halo, i.e., its inner flattened and outer spherical
density distribution, is reproduced, at least qualitatively, by the present
model. The outer spherical halo is formed via essentially dissipationless
merging of small subgalactic clumps, whereas the inner flattened one is formed
via three different mechanisms, i.e., dissipative merging between larger,
more massive clumps, adiabatic contraction due to the growing Galactic disk,
and gaseous accretion onto the equatorial plane.

(4) For the simulated metal-poor stars with [Fe/H] $\le$ $-1.0$,
there is no strong correlation between metal abundances and orbital
eccentricities, in good agreement with the recent observations.
Moreover, the observed fraction of the low-eccentricity stars
is reproduced correctly for [Fe/H] $\le$ $-1.6$ and approximately
for the intermediate abundance range of $-1.6$ $<$ [Fe/H] $\le$ $-1.0$.

(5) The mean rotational velocity of the simulated halo, $<V_\phi>$, is
somewhat positive (prograde) at [Fe/H] $<$ $-2.2$ and increases linearly with
[Fe/H] at [Fe/H] $>$ $-2.2$. The stars at smaller distance from the disk
plane appear to show systematically larger $<V_{\phi}>$. 

Based on these results, we discuss how early processes of dissipationless and
dissipative merging of subgalactic clumps can reproduce plausibly and
consistently the recent observational results on the Galactic stellar
halo. We also present a possible scenario for the formation of the entire
Galaxy structure, including bulge and disk components, in conjunction with
the halo formation.

\end{abstract}


\section{Introduction}

Formation and evolution of the metal-deficient halo remain elusive,
although it constitutes a fundamental component in the Galaxy. Its detailed
structure and kinematics are clues to the understanding of how the
Galaxy has developed its dynamical structures (e.g., Freeman 1987;
Majewski 1993; van den Bergh 1996). 
In particular, observed correlation between chemical and dynamical properties
of halo stars or lack thereof provides valuable information on the early
dynamical history of the Galaxy, because the metallicity can be regarded as
a sort of ``clock'' of the Galaxy  (Gilmore, King, \& van der Kruit 1990).
Also, studies of kinematic substructures in the stellar
halo allow us to elucidate the detailed  history of merging between the Galaxy
and any subgalactic fragments, such as dwarf galaxies, because the time scale
for mixing of halo stars due to momentum and energy exchange
exceeds the age of the Galaxy (e.g., Lynden-Bell \& Lynden-Bell 1995;
Johnston, Spergel, \& Hernquist 1995; Helmi et al. 1999; Morrison et al. 2000).

Based on penetrative analysis of such important populations
of the Galactic ``fossil record'',
two canonical scenarios of the Galactic halo formation were proposed,
that have long been influential for later observational
and theoretical studies of disk and elliptical galaxies.
One is the monolithic collapse picture proposed
by Eggen, Lynden-Bell, \& Sandage (1962, hereafter ELS),
in which the halo is formed from an overdense homogeneous
sphere very rapidly within an order of free-fall time  ($\sim$ $10^8$ yr).
The other is the accretion/merging scenario by
Searle \& Zinn (1978, hereafter SZ),
in which  the halo is formed slowly 
($\sim$ $10^{9}$ yr) by chaotic merging/accretion of
several subgalactic fragments.
Although a considerable number of works have
discussed advantages and disadvantages of these scenarios 
in explaining accumulated observational results 
(e.g., Yoshii \& Saio 1979; Norris, Bessell,
\& Pickles 1985; Norris \& Ryan 1991; Beers \& Sommer-Larsen 1995),
it is yet unsettled which picture, ELS or SZ,
can describe more correctly and plausibly 
the early dynamical evolution the Galaxy (e.g., van den Bergh 1996).

In order to settle the above issue, we require a large and reliable
set of data for metal-poor halo stars chosen with criteria that do not unduly
influence the subsequent analysis (e.g., Norris 1986; Chiba \& Yoshii 1998,
hereafter CY). In particular, the large data set of halo stars with
various metallicities chosen without any kinematical selection bias are
required for reliable discussion on the issue. Such kinematic aspects
of the Galactic metal-poor stars have been greatly improved
by the recently completed {\it Hipparcos} Catalogue (ESA 1997) and various
ground-based catalogs (e.g., Platais et al. 1998; Urban et al. 1998) which
provide unprecedentedly accurate proper motion data for a wealth of metal-poor
stars (Beers et al. 2000).
Recent works using non-kinematically selected samples of stars 
with  available proper motions revisited the issue
and demonstrated that either of the two scenarios alone
has several difficulties in explaining the up-to-date observational results 
(Chiba \& Beers 2000, hereafter CB).

For example, the ELS scenario is inconsistent with 
the observed no correlation between metallicities
and orbital eccentricities of halo stars (CY; CB), and
also the lack of the halo abundance gradient (CY).
Equally, the SZ scenario unlikely explains
the observed vertical gradient of rotational velocity
as well as the inner flattened density distribution of halo stars (CB).
This failure of the two models suggests several researchers to consider
alternative models  such as a hybrid scenario of the halo
formation in which dissipationless merging of subgalactic clumps forms 
the outer Galactic halo, whereas ordered dissipative collapse
forms the inner one (e.g.,  Norris 1994; Freeman 1996; Carney et al. 1996). 
However, while such a hybrid scenario may reproduce
most of observational results, it is indispensable for the scenario 
to explain whether both of these dissipationless and dissipative processes
actually take place in the early dynamical evolution of the Galaxy.

Besides these two canonical scenarios, the formation of the stellar halo
has not been discussed so extensively in the framework of
the currently favored theory based on hierarchical assembly of cold dark matter
(CDM) halos (e.g., White \& Rees 1978; Peacock 1999).
In the hierarchical clustering scenario,  as the gravitational collapse
of a protogalaxy proceeds, numerous subgalactic clumps are developed 
from initial small-scale density perturbations predicted by the CDM model. 
These clumps merge with one another dissipatively or dissipationlessly
to  grow in a hierarchical manner and finally form the Galaxy. 
Recent numerical simulations of the Galaxy formation based on
the hierarchical clustering scenario demonstrated that metal-poor stars
which finally build the stellar halo component have already been formed
at very high redshift ($z$ $>$ 2) within each of merging subgalactic clumps
(Steinmetz \& M\"uller 1995; Bekki \& Chiba 2000a).  
These studies also suggested that the dispersal of metal-poor stars caused by
tidal disruption of clumps is responsible for the formation of the
stellar halo.

Previous simulations conducted to date have provided a rough sketch of
the Galaxy formation and demonstrated the importance of initial small density
perturbations in the formation processes of galactic bulges and disks (e.g.,
Katz 1992). Furthermore semi-analytic models of galaxy formation and evolution
based on CDM models have provided important predictions on bulge and disk
formation (e.g., Kauffmann et al. 1993; Baugh et al. 1998; Somerville \&
Primack 1999). However, it was yet unexplored whether or not the CDM model
can provide a definite and realistic picture for the formation of the Galactic
stellar halo, including all of its structural, kinematical, and chemical
properties.

Thus, the purpose of this paper is (1) to investigate thoroughly whether 
the CDM model can reproduce successfully the recent observational results on
the Galactic halo and (2) to provide a general picture of the Galaxy formation
including stellar halo, bulge, and disk components. We perform  numerical
simulations of the Galaxy based on the standard CDM model and investigate
structure and kinematics of the developed metal-poor stellar halo.
In particular, we compare our numerical simulations with recent
observational results by CY and CB and elucidate the detailed mechanism of
the halo formation.

We note that C\^ote et al. (2000) recently investigated the origin of 
metallicity distribution of the Galactic stellar halo in the context
of hierarchical merging of dwarf-like galaxies. However, the detailed
kinematical properties of the halo were unexplored. Also,
most of the CDM-based numerical models on disk galaxy formation have focused
on only the fundamental properties of a disk, such as an exponential density
profile (Katz 1992) and Tully-Fisher relation (Steinmetz \& Navarro 1999).
The spatial structure of the stellar halo has been examined by Steinmetz \&
M\"uller (1995), but, again, the detailed kinematics of the halo stars in the
simulated model remained unknown.
Thus we believe that our approach to modeling all of the
structural, kinematical, and chemical properties of the halo
can provide valuable clues to the formation of the Galaxy.

The layout of this paper is as follows.
In \S 2, we summarize  numerical models used in the
present study and describe the methods for 
analyzing structure and kinematics of the simulated stellar halo.
In \S 3, we present numerical results on the time evolution of 
morphology,  metallicity distribution,  and dynamical properties of the 
forming halo. 
In \S 4, we discuss success and failure of the present
model in reproducing the observed dynamical properties of the Galactic
halo and provide a possible scenario of the Galaxy formation.  
The conclusions of the preset study
are  given in \S 5. 

\section{Model}

\subsection{Numerical methods}

The numerical method and technique for solving galactic chemodynamical
evolution and models for describing star formation and dissipative gas dynamics
are presented in Bekki \& Shioya (1998), so here we briefly describe the
initial conditions of protogalactic clouds, star formation law, and chemical
evolution model. The way to set up initial conditions for numerical simulations
of forming disk galaxy within a hierarchical clustering scenario is essentially
the same as that adopted by Katz \& Gunn (1991) and Steinmetz \& M\"uller
(1995). 
We here use the COSMICS (Cosmological Initial Conditions and
Microwave Anisotropy Codes), which is a package
of fortran programs for generating Gaussian random initial
conditions for nonlinear structure formation simulations
(Bertschinger 1995; Ma \& Bertschinger 1995).
We consider an isolated homogeneous, rigidly rotating sphere, on which
small-scale fluctuations according to a CDM power spectrum are superimposed.
The mass ratio of dark matter to baryons is 9.0
and both components have the same initial distribution. 
The initial total mass, radius, and spin parameter ($\lambda$) of the sphere
in the present model are $6.0\times10^{11}\rm M_{\odot}$, 45 kpc, and
0.08, respectively. For cosmological parameters of $\Omega =1.0$, $q_0=0.5$,
and $H_{0}=50$ km $\rm s^{-1}$ ${\rm Mpc}^{-1}$ adopted in the present study,
the initial overdensity ${\delta}_{i}$ of the sphere in the fiducial model is
estimated to be 0.29 corresponding to a 2.5 $\sigma$ overdensity
of a biased CDM spectrum on a mass-scale of 6.0 $\times 10^{11} \rm M_{\odot}$ 
with a biasing parameter $b = 2$.
We start the simulation at the redshift $z = 25$ and follow it 
till $z=0$  corresponding to the age of the universe equal to  13 Gyr.
Initial conditions similar to those adopted in the present study
are demonstrated to be plausible and realistic for the formation
of the Galaxy (e.g., Steinmetz \& M\"uller 1995).

Star formation is modeled by converting the collisional gas particles into
collisionless new stellar particles according to the Schmidt law (Schmidt 1959)
with the exponent of 2 and the coefficients in the law are taken from
the work of Bekki (1998).
The collisional and dissipative nature of the interstellar medium are modeled
according to the sticky particle method (Schwarz 1981)
with the cloud radius ($r_{cl}$) of 450 pc and we consider multiple collisions
among clouds (see Bekki \& Shioya 1998 for details). 
It should be emphasized here that this discrete cloud model can at best
represent the {\it real} interstellar medium of galaxies in a schematic way.
As is modeled by McKee \& Ostriker (1977), the interstellar medium is
composed mainly of `hot', `warm', and `cool' gas, each of which mutually
interacts hydrodynamically in a rather complicated way.
Actually, such a considerably complicated nature of the interstellar medium
in forming disk galaxies would not be so simply represented by the ``sticky
particle'' model in which gaseous dissipation is modeled by ad hoc
cloud-cloud collision: Any existing numerical method probably could
not model the {\it real} interstellar medium in an admittedly proper way.
In the present study, as a compromise,
we only try to address some important aspects of hydrodynamical
interaction between interstellar clouds in forming disk galaxies.
More elaborated numerical modeling of the interstellar medium
would be necessary for our further understanding of dynamical evolution
in forming disk galaxies.

We note that the phenomenological sticky particle model adopted here is
in great contrast with SPH models used in previous numerical works of
galaxy formation (e.g., Katz 1992; Steinmetz \& M\"uller 1995).
Although the sticky particle model has succeeded in addressing 
some important aspects of hydrodynamical interaction in interstellar medium,
such as dynamical interaction between giant molecular clouds 
(e.g., Roberts \& Hausman 1984; Combes \& Gerin 1985),
its weak point is considered to be an ad hoc estimate
of an energy dissipation in cloud collisions (e.g., Shlosman \& Noguchi 1993).
On the other hand, the SPH method, which often introduces the virtual cut-off
of  cooling function around $10^4$ $K$ (e.g., Katz 1992) and thus can
investigate only warm gas components ($T$ $>$  $10^3$ -- $10^4$ $K$),
cannot follow formation and evolution of cold molecular components that may
well dominate the interstellar medium in very gas-rich young disk galaxies.
Accordingly the SPH method holds difficulties in simulating interaction
between {\it discrete} gas clouds in a self-consistent manner.
Thus, we consider that both the SPH and sticky particle models
are complimentary with each other in representing the real interstellar
medium of forming galaxies.

Chemical enrichment through star formation during merging of subgalactic clumps
is assumed to proceed both locally and instantaneously in the present study.
The model for analyzing metal enrichment of each gas and stellar particle
is as follows.
First, as soon as a gas particle is converted into a new stellar one by
star formation, we search neighboring gas particles locating within
0.01 in our units (corresponding to 450pc) from the position of the new
stellar particle and then count
the number of neighboring gas particles, $N_{\rm gas}$.
Next, we assign the metallicity of the original gas particle to the new
stellar particle and increase the metal of each neighboring gas particle
according to the following equation about the chemical enrichment:
  \begin{equation}
  \Delta M_{\rm Z} = \{ Z_{i}R_{\rm met}m_{\rm s}+(1.0-R_{\rm met})
 (1.0-Z_{i})m_{\rm s}y_{\rm met} \}/N_{\rm gas} 
  \end{equation}
where $\Delta M_{\rm Z}$ represents the increase of metal for each
gas particle. $Z_{i}$, $R_{\rm met}$, $m_{\rm s}$, and $y_{\rm met}$ represent
the metallicity of the new stellar particle (or that of the original gas
particle), the fraction of gas returned to the interstellar medium,  the
mass of the new star, and the chemical yield, respectively.
The values of $R_{\rm met}$ and $y_{\rm met}$ are set to
be 0.3 and 0.02, respectively.

The total particle number used for modeling the initial sphere is 14147 both
for dark matter and for baryons (gas and new stars), 
which means that the mass of each particle is
3.8 $\times$ $10^7$ $ \rm M_{\odot}$ 
for dark matter and 4.2 $\times$ $10^6$ $ \rm M_{\odot}$ for baryons.
All the calculations related to the above chemodynamical evolution
have been carried out on the GRAPE
board (Sugimoto et al. 1990)  at Astronomical Institute of Tohoku University.
The parameter of gravitational softening is set to be fixed at 0.053 in our
units (2.4 kpc). The time integration of
the equation of motion is performed by using the 2nd-order leap-flog method.

As Figure 1 reveals, showing the evolution of dark matter particles, our
simulation does not allow the persistence of small-scale structures at $z=0$,
as in Katz (1992) and Steinmetz \& M\"uller (1995) using similar particle
numbers, but in sharp contrast to recent high-resolution N-body studies
(Klypin et al. 1999; Moore et al. 1999). These authors resolved the detailed
clustering process of dark halo clumps with masses down to the order of
$\sim$ $10^8$ $M_\odot$. The reasons for this difference are that the
total particle number in the present study ($N\sim$ a few $\times 10^4$) is
considerably smaller than those adopted by such high-resolution simulations
($N\sim$ $10^6$ or mesh number of $128^3$), and that the rather large length
of gravitational softening (2.4 kpc) in our study erases small-scale
structures with masses of $10^7\sim10^8$ $M_{\odot}$
even if they are developed. While it is yet unknown how such small-scale
structures are modified by inclusion of gas dynamics and star formation,
our primary concern here is to figure out the general formation process of
the stellar halo in the framework of clustering theory, without going into
the fine details of the halo structure. Perhaps, any additional phenomena
associated with masses of $10^7\sim10^8$ $M_{\odot}$, such as excessive
substructures of the halo and/or dynamical heating of the disk, may be
precluded in a class of CDM theory (Kamionkowski \& Liddle 1999;
Spergel \& Steinhardt 2000).

\subsection{Main points of our analysis}

Based on the above model, we mainly investigate structural and kinematical
properties of the simulated metal-poor component:
the mass fraction of the simulated metal-poor stars with [Fe/H] $\le$ $-1.6$,
compared with the total baryonic mass of the system, is 0.014 at $z=0$ and
their mean age is about 10.5 Gyr. We focus on the nature of these
old and metal-poor stars based on the present simulations, to investigate
whether or not the CDM theory can reproduce reasonably well the observed
fundamental characteristics of the Galactic halo.
We investigate the following five points: (1) How does each
of the Galactic components, halo, bulge, and disk, form during merging between
subgalactic clumps and what is the mutual relationship among the
formation processes of these components?,
(2) How is the inner flattened and outer spherical halo developed?,
(3) Can the CDM model reproduce the observed dependence of $<V_{\phi}>$ on
[Fe/H]?,  
(4) What are the characteristics of  the velocity ellipsoid of
the simulated halo?, 
and (5) Are there any kinematic substructures in  the simulated halo?

As a part of out analysis described in the next section, we calculate
orbital eccentricities, $e$, of the stars with [Fe/H]$\le-0.6$ at the
epoch $z = 0$. Here $e$ for each stellar particle is defined as:
\begin{equation}
  e= \frac{r_{apo}-r_{peri}}{r_{apo}+r_{peri}} \;
\end{equation}
where $r_{apo}$ and $r_{peri}$ are apo-galactic and peri-galactic distances
from the center of the simulated Galaxy, respectively.
For estimating $e$, we first select the stellar particles with [Fe/H]$\le-0.6$
found at $z=0$. Then we calculate the time evolution of their orbits
under the gravitational potential of the simulated disk Galaxy achieved at
$z = 0$ for ten dynamical time scale ($\sim$ 1.8 Gyr), and then estimate $e$.
In order to avoid the contamination of metal-poor bulge stars in this
estimation, we select only particles with their apo-galactic distances
ranging from 8.5 to 17.5 kpc.

We use the symbol [Fe/H] as the total metal abundance
instead of the symbol $Z$, to avoid confusion with distance from the disk
plane, although the current model does not consider the evolution of each
element separately.

\placefigure{fig-1}
\placefigure{fig-2}
\placefigure{fig-3}
\placefigure{fig-4}
\placefigure{fig-5}

\section{Result}

\subsection{Formation process of the entire Galaxy structure}

We first describe the formation process of the entire Galaxy structure,
consisting of dark matter halo, stellar halo, bulge, and disk components.
Figure 1, 2, and 3 show dynamical evolution of dark matter, gas, and
stars formed from gas, respectively, in the present model. The turn-around
and collapse redshifts of the protogalactic sphere are 3.25 and 1.68,
respectively. As the sphere expands, numerous small clumps composed of
dark matter develop from non-linear gravitational growth of
initial local density maxima at high redshift, $z$ $\sim$ 10.
Owing to the strong gravitational attraction of dark matter, 
gas particles also  assemble  into each of the  clumps. Then,
gaseous dissipation due to collisions between gas particles proceeds,
which results in the central concentration of gas
even in this early dynamical stage (5 $<$ $z$ $<$ 10).
Consequently, first stars are formed in the central part of each clump.

Firstly born stars
with old ages ($>$ 10 Gyr) and low metallicities ([Fe/H]$<-3$) are confined
within these small clumps until they are disrupted by later mutual merging.
The clumps continue to merge with one another and grow up to become
more massive clumps ($z$ $=$ 2.56, 4.47).   
Star formation proceeds efficiently in these growing clumps
($z$ $=$ 1.75), and consequently two most massive clumps are formed after
multiple merging events of small clumps ($z$ $=$ 1.75). 
These two clumps dissipatively merge with each other and then leave a central
bulge-like component ($z$ $=$ 1.28).
Other surviving, less massive clumps continue to merge with one another and
also with the two massive clumps even after the strong merging event of
these massive clumps.
Gas falling onto the surroundings of the bulge forms a disk-like
component composed of gas and gradually forming stars
(0.75 $<$ $z$ $<$ 0.94).
This young, disk-like component grows gradually owing to the continuous 
gas accretion from the halo region and finally becomes a well-developed disk
with its size and mass similar to those of typical present-day disk
galaxies like our own (0 $<$ $z$ $<$ 0.45).  
This chemodynamical evolution of the Galaxy is basically
in agreement with earlier numerical results by Katz (1992) and
Steinmetz \& M\"uller (1995).

Figure 4 shows the star formation history of the present model.
While the Universe is still young, of the age of 0.3--0.4 Gyr
($z$ $\sim$ 10), some gas inside subgalactic clumps developed from small-scale
density maxima is already consumed to form stars at a small
rate of $\sim$ 2  $M_{\odot}$ ${\rm yr}^{-1}$.
These very old and metal-poor stars may well be regarded as the first stars
in the Galaxy\footnote{Many of these first stars are too metal-poor, with
[Fe/H]$<-4$, which have not yet been found by extensive surveys of such stars
to date (Beers 2000). This problem may be precluded if we consider the
pre-enrichment of gas in a cosmological scale (e.g., Cen \& Ostriker 1999).}.
Star formation rate of the forming Galaxy exceeds
5 $M_{\odot}$ ${\rm yr}^{-1}$ at 1.5 $<$ $z$ $<$ 2.0 (corresponding to
$\sim$ 3 Gyr after the onset of the collapse)
and takes the maximum value of $\sim$ 16 $M_{\odot}$ ${\rm yr}^{-1}$ when
the two massive clumps merge with each other ($z$ $\sim$  1.5).
It is worthwhile here to remark that this starburst epoch of $z$ $\sim$ 1.5
matches the formation epoch of the bulge-like component in the simulated
Galaxy. This starburst also
drives rapid chemical evolution in the central part of the Galaxy
and thereby greatly enriches gas and stars. As a consequence,
this bulge-like component is made very metal rich.

After the star formation rate reaches the peak value, it gradually, at a
nearly constant rate, declines to a few $M_{\odot}$ ${\rm yr}^{-1}$,
because minor merging which triggers sporadic increase of star formation
does not frequently occur after $z$ $=$ 1.28.
Stars formed from this gradual and long-term star formation build up
the present-day disk component in our model.
In this regard, Blitz et al. (1999) argued that the so-called High Velocity
Clouds (HVCs) having large negative radial velocities may correspond to
external primordial gas which continues to accrete toward the Galactic disk.
If this process, which is not included in our simulation, is at work,
HVCs would modify the long-term star formation history of the
simulated disk at 0 $\le$ $z$ $<$ 1.

Figure 5 shows the relation between ages and metallicities of the stars
formed in the simulated Galaxy. 
As a natural result of chemical evolution inside clumps,
more metal-rich stars tend to have younger ages.
The excess of stars can be seen in the age between 9  and 10 Gyr,
which corresponds to the epoch of starburst triggered by merging between
two massive clumps at  $z$ $\sim$ 1.5.
Also, there is a large dispersion in metallicities for stars with given ages
(e.g., metallicities range from $-1.0$ to 0.4 dex for stars with ages
$\sim$ 5 Gyr), in sharp contrast to the prediction of a simple, one-zone
chemical evolution model. This dispersion in the relation between
metallicities and ages is explained in the following way.
First, chemical evolution proceeds independently in each clump
having mutually different gaseous metallicity, 
so that the stars born in different clumps have different metallicities,
even if they are born at the same epoch. Second, in contrast to a one-zone
model, mixing of metals is incomplete, so that gaseous metallicity is made
considerably inhomogeneous inside each clump.

The present numerical simulation as well as other chemodynamical models
based on the standard CDM model (Steinmetz \& M\"uller 1995; Bekki 2000;
Bekki \& Chiba 2000a) reveal that the number fraction of
metal-poor disk stars with metallicities less than 0.25 $Z_{\odot}$
(where  $Z_{\odot}$ is solar metallicity) is estimated to be 10 -- 20 \%,
which is still inconsistent with the observed value of 2 \%. Thus, these
CDM-based models encounter the so-called G-dwarf problem
in the solar neighborhood (e.g., Thuan et al.  1975). We note that
this failure is due principally to the adopted initial
condition (i.e., isolated gaseous sphere) and that the problem may be solved
by invoking the later accretion of gas-rich HVCs.
The physical process related to the accretion of HVCs onto
the Galactic disk and its role in solving
the G-dwarf problem will be presented elsewhere.

Figure 6 and 7 present the final spatial distribution of the stars at
$z$ $=$ 0, for a given metallicity and age range, respectively. It is clear
that the distribution of more metal-poor (metal-rich) stars is likely to be
more spherical (disky), which means that progenitor gas clouds for more
metal-poor (metal-rich) stars experience less (more) amount of gaseous
dissipation in the early dynamical evolution.
Furthermore the distribution of older (younger) stars is likely to be
more spherical (disky), which is a consequence of
the process that old bulge and halo components are formed basically
through merging between subgalactic clumps, 
whereas a young disk is formed from later gradual and dissipative
gaseous accretion.
It is also clear from these figures that the central bulge component contains
old and metal-rich stars, though it emerges at moderate redshift
($z$ $\sim$ 1.5) after the major merging of two massive clumps:
in the present hierarchical picture, a significant
fraction of stars that compose the present-day bulge are already formed
at high redshifts ($z$ $>$ 2) within the central region of clumps.
Thus, this suggests us that even if
the Galactic bulge is observed to contain old and metal-rich stars
(McWilliam \& Rich 1994), it does {\it not} necessarily indicate that the
bulge is formed {\it in} {\it situ} at high redshift. This aspect for the
bulge formation was also emphasized by Noguchi (1999).

The  disk component, on the other hand, is basically young and metal-rich
compared with other spherical components, because the disk is formed by later
accretion of enriched gas, which is tidally stripped from subgalactic clumps
during their merging event. Furthermore the disk thickness is systematically
smaller for more metal-rich stars, which suggests that gaseous dissipation
plays an important role in determining the disk thickness. It is noted that
the thickening of the forming disk is also driven by two-body scattering
of disk stars by their interaction with rapidly moving high-mass dark matter
particles (e.g., Steinmetz \& M\"uller 1995).
Here we are unable to separate between the so-called thick and thin disks owing
to the resolution of the present simulation ($\sim$ 2 kpc, which is
comparable to or larger than the characteristic thickness of both disk
components). 
However, {\it if} the thicker disk obtained from our simulation
with the age between 8 and 10 Gyr
(corresponding to the lower middle panel in Figure 7) corresponds to
the Galactic thick disk,
Figure 7 implies  that it is formed while surviving less massive
clumps are still merging with the young Galaxy at the epoch of
8 -- 10 Gyr. We also note that, from the lower middle panel of
Figure 7, the morphology of the early Galaxy about 8 Gyr ago may have
looked like a S0 galaxy or a disky elliptical galaxy, rather than
currently seen -- we here stress that not all of stars with the age
between 8 -- 10 Gyr have already assembled into the Galaxy 8 Gyr ago.
This implies that some high redshift galaxies, morphologically
classified as S0s, can be transformed to more late-type
disk galaxies at lower redshift,
owing to later gaseous infall and the resultant disk growth.

\placefigure{fig-6}
\placefigure{fig-7}
\placefigure{fig-8}

\subsection{Halo properties}

In the following, we focus on the results for the metal-poor stars with
[Fe/H] $\le$ $-1.0$ that are expected to build up the present-day
Galactic stellar halo.

\subsubsection{Star formation history}

Figure 8 shows the number fraction of the stars having older ages than a given
age (cumulative age distribution), for the metallicity ranges of
[Fe/H] $\le$ $-1.6$ (solid line) and $-1.6$ $<$ [Fe/H] $\le$ $-1.0$
(dotted line), respectively. In other words, the plot shows the number fraction
of the stars which have already been form by the given epoch.
Since these metal-poor stars are considered to compose the present-day
Galactic halo, this figure is expected to describe its star formation
history. It follows that most of metal-poor stars ($\sim$ 80\%) with
[Fe/H] $\le$ $-1.6$ are already formed 10 Gyr ago ($z$ $>$ 2),
thereby confirming that 
such stars are born initially inside subgalactic clumps at high redshifts.
The figure also suggests that the time scale for the formation of old,
metal-poor stars is of the order of a few Gyr, not $10^8$ yr
as was envisaged by ELS.

On the other hand, the stars with $-1.6$ $<$ [Fe/H] $\le$ $-1.0$ still
continuously form between 6 Gyr and 10 Gyr, though a significant fraction
of them (42 \%) have been already formed at $z$ $>$ 2. 
These relatively young, more metal-rich halo stars are formed from dissipative
accretion of already enriched gas that was initially within subgalactic
clumps but later stripped tidally during their merging event.
This result suggests that some metal-poor halo stars with
$-1.6$ $<$ [Fe/H] $\le$ $-1.0$ show younger ages than those with
[Fe/H] $\le$ $-1.6$. These young populations might show small abundance
ratios such as [Mg/Fe], compared with old populations with [Fe/H] $\le$ $-1.6$.
It is interesting to explore if future observations
reveal the systematic age difference
between [Fe/H] $\le$ $-1.6$ and $-1.6$ $<$ [Fe/H] $\le$ $-1.0$.

\subsubsection{Morphology and structure}

Figure 9 shows how the metal-poor halo populations with [Fe/H] $\le$ $-1.6$
are formed during the collapse of the Galaxy. Clearly, a large fraction of such
stars ($\sim$ 50 \%) have already been formed inside small-scale density
maxima at early epoch $z \ga 2.6$. Then, the numerous small clumps, developed
from these density maxima, merge or tidally interact with one another, and
stars inside them are spread over the outer part of the halo. We note that
these early
processes proceed basically in a dissipationless way, because gas inside
small clumps are quickly consumed by star formation or disrupted by tidal
interaction. While these events take place, two massive clumps are
developed by their gravity and spiral toward the inner part of the halo.
These massive clumps, composed of both stars and gas, are then
tidally disrupted during their strong interaction and merging
($z$ $=$ 1.75, and 1.28). Many metal-poor stars stripped by
these processes are spread over the inner part of the forming Galaxy and
build up the dense, inner part of the halo. To summarize, the halo formation
involves dissipationless merging of many small clumps in its outer part and
dissipative merging of two massive clumps in its inner part.

As is shown in Figure 10, the radial density distribution of the halo at
$r \le 20$ kpc is similar to the observed power-law,
$\rho(r) \propto r^{-3.5}$, where $r$ is the distance from the center of
the disk (Saha 1985; Hawkins 1984; Preston et al. 1991; Sommer-Larsen \& Zhen
1990; CB). Figure 11 and 12 show the meridional distributions of
the stars with [Fe/H] $\le$ $-1.6$ (left panels) and
$-1.6$ $<$ [Fe/H] $\le$ $-1.0$ (right panels) at two different redshifts,
$z$ = 0 (a) and 0.97 (b). 
It follows from the panels for $z$ $=$ 0 that the inner part of the halo
is appreciably flattened, in particular, for stars with 
$-1.6$ $<$ [Fe/H] $\le$ $-1.0$, whereas the outer part is more likely
to be spherical. Also, we see the lack of stars along the polar axis.
These simulated results are in good agreement with the observed dual
nature of the stellar halo, characterized by an inner flattened part
at $R \la 15$ kpc and an outer spherical part (Hartwick 1987; 
Sommer-Larsen \& Zhen 1990; Preston et al. 1991; Kinman et al.
1994; Layden 1995; CB).
We note that this dual nature explains why studies of star counts
generally yield an approximately spherical halo, whereas the local
anisotropic velocities of the halo stars suggest a highly
flattened system (Freeman 1987).

We now describe, in more detail, probable three physical processes involved in
this duality of the halo.

The first is the difference in dynamical evolution of subgalactic clumps
with different  masses. Less massive and smaller clumps, that are formed from
initial small-scale density fluctuations and do not  grow steadily  by merging
with other clumps, are rather susceptible to destruction by strong tidal
interaction and merging in the course of gravitational collapse of the Galaxy.
The stars born inside such clumps are spread over the outer region of
the proto-Galaxy after their destruction.
Also, these less massive clumps are more likely to be completely destroyed
{\it before} they spiral toward the central region of the proto-Galaxy,
so that the tidally stripped stars are located in the outer region of
the halo. Furthermore, since the tidal stripping occurs randomly among
the clumps with variously different orbits, the disrupted stars
also have mutually different orbits.
This leads to the outer halo with spherical shape and with
no systematic rotation, which is in good agreement with
the reported properties of the outer halo (CB).

On the other hand, more massive clumps, that grow rapidly
owing to multiple dissipative merging with other smaller clumps,
gradually move toward the central region of the proto-Galaxy because
of dynamical friction. These clumps are less likely to be completely
destroyed by interaction with other small clumps. Since dynamical friction
between clumps and background dark halo tends to circularize  their orbits,
they have a large amount of orbital angular momentum. Furthermore this
dynamical friction makes the clumps approach to the Galactic plane
(defined in the present epoch) before their merging. Finally,
these massive clumps merge with each other in a considerably dissipative way
owing to the large fraction of gas, and are consequently
destroyed by the violent relaxation.
Since most of the stars stripped from the clumps have orbital motions
with a finite angular momentum, similarly to the clumps themselves,
the aftermath of this merging
event shows a flattened halo with finite, prograde rotation.
Thus, the present numerical model based on the standard CDM model
suggests that both dissipationless merging in the outer part
and dissipative merging in the inner part
are responsible for the observed dual nature of the halo.

The second key process, which is only important for the inner halo,
is the dynamical effect of the gradually growing disk on the halo
structure (Binney \& May 1986; Chiba \& Beers 2001).
As is shown in Figure 12, the spatial distribution of the stars with
$-1.6$ $<$ [Fe/H] $\le$ $-1.0$ for $R$ $<$ 15 kpc is more flattened at
$z$ $=$ 0 than at $z$ $=$ 0.97. As the disk is not so developed  at
$z$ $=$ 0.97 (see Figure 2 and 3), this result suggests that the growth of
the disk component due to later gaseous accretion compresses adiabatically
the halo with an initially less flattened distribution, thereby leading
to a more flattened halo in its inner part.
This effect of adiabatic compression can be more clearly seen in the
more metal-rich stars with $-1.6$ $<$ [Fe/H] $\le$ $-1.0$ than
[Fe/H] $\le$ $-1.6$, possibly because an ensemble of more metal-rich stars
is systematically `colder' owing to more dissipation of gas, before
the disk growth.

The third process, which is also responsible for the formation of the
inner halo alone, is drawn by comparing the spatial distributions of stars
at the epoch of $z$ $=$ 0 with $z$ $=$ 0.97 in Figure 11.
It appears that in the range of $10 \la R \la 20$ kpc near the plane,
there are a larger number of stars with $-1.6$ $<$ [Fe/H] $\le$ $-1.0$
at $z$ = 0 than at $z$ = 0.97.  Furthermore, these stars at $z$ = 0 clearly
have a highly flattened distribution.
These results suggest that such stars located in the range of
$10 \la R \la 20$ kpc at the current epoch have been formed since $z$ $=$ 0.97
and that these newly born stars build up a flattened density distribution.
The probable reason for its flattened distribution is that these stars are
formed from later gaseous accretion onto the outer part of the disk
and thus sustain a large amount of gaseous dissipation prior
to their birth, as was envisaged by Chiba \& Beers (2001).
Such  later formation of a highly flattened halo composed mostly of relatively
young (8--10 Gyr) stars cannot be seen clearly for the more
metal-poor stars with [Fe/H] $\le$ $-1.6$. This is consistent with
the fact that most of stars with [Fe/H] $\le$ $-1.6$ are formed more than
10 Gyr ago (see Figure 8).
Thus, we can conclude that the outer nearly spherical halo is
formed by merging of less massive subgalactic clumps,
whereas the inner flattened halo is formed by three different mechanisms:
highly dissipative merging between more massive clumps, adiabatic compression
due to secular and gradual disk growth, and later dissipative
gaseous accretion onto the disk plane.

It should also be remarked that in all metallicity ranges and redshifts,
there appears the lack of stars along the polar axis of the simulated halo.
This was actually obtained from kinematic analyses of nearby metal-poor stars
(Sommer-Larsen \& Zhen 1990; CB), but was attributed to the small probability
that stars exploring near the polar axis are represented in the solar
neighborhood. However, our model suggests that it is {\it real}.
It is possibly due to the effect of dynamical friction between dark matter
halo and subgalactic clumps: dynamical friction makes clumps approach to
the disk plane, so that the spatial distribution of metal-poor stars,
which are tidally stripped from these clumps, tend to avoid the region
near the polar axis.

\placefigure{fig-9}
\placefigure{fig-10}
\placefigure{fig-11}
\placefigure{fig-12}

\subsubsection{[Fe/H] vs $e$ relation}

Figure 13 shows that there is no significant correlation between [Fe/H] and $e$
for the stars with [Fe/H]$\le-0.6$, and that the existence of low eccentricity
($e<0.4$), low metallicity ([Fe/H] $<-1$) stars is successfully reproduced
in the present CDM model. This may be explained in a following manner.
First, as ELS argued, the rapid contraction of a gravitational potential within
a dynamical time ($\sim$ $10^{8}$ yr) results in the transformation of
initially nearly circular (smaller $e$) orbits to very eccentric (larger $e$)
ones. On the other hand, the eccentricities of the orbits remain basically
unchanged if the contraction is slow enough ($\sim$ $10^{9}$ yr). In the
present CDM model, the time scale for the contraction of the Galaxy
is lengthened by the expanding background universe
(the collapse time scale of a proto-galactic
sphere with a turn-around radius of $\sim100$ kpc is of the order of Gyr),
 so that the process of star formation mainly
triggered by merging of small clumps is rather extended ($\sim$ 2 Gyr). Thus,
the orbital eccentricities of the metal-poor stars, once formed, are not
greatly influenced by the change of an overall gravitational potential of the
Galaxy.
Second, as we mentioned above, most of the metal-poor stars have been confined
within the massive clumps, where their orbits are gradually circularized due to
dissipative merging with smaller clumps and dynamical friction with the
background dark halo. Thus, a finite fraction of the debris stars after the
last merging event preserve the orbital angular momentum of the clumps.
Both of these processes may give rise to the existence of low-$e$, low-[Fe/H]
stars in the simulated Galactic halo. 
Although the  observational evidence of no significant correlation between
metallicity and eccentricity in metal-poor halo stars of the Galaxy is
suggested to conflict with ELS's monolithic scenario (CY), it has not been
explicitly clarified whether this observational evidence is consistent with
the merger picture that is originally proposed by SZ.
Thus the present numerical study first demonstrates that
the SZ picture is more plausible to explain the observed [Fe/H]-$e$ relation.

To be more quantitative, we plot, in Figure 14, the cumulative $e$ 
distributions of the metal-poor stars with [Fe/H]$\le-1.6$ (solid line) and
$-1.6<$[Fe/H]$\le-1.0$ (dotted line). For the halo component with
[Fe/H]$\le-1.6$, the fraction of the simulated low-$e$ stars with $e<0.4$ is
about 0.17, which is in good agreement with the observational result of about
0.2 (CY; CB). Also, as is consistent with the observational result, the
cumulative $e$ distribution in the intermediate abundance range
$-1.6<$[Fe/H]$\le-1.0$ is systematically larger than that for [Fe/H]$\le-1.6$.
The fraction of the simulated low-$e$ stars with $e<0.4$ in this abundance
range ($\sim 0.45$) is somewhat larger than the observation ($\sim 0.35$),
suggesting that the metal-weak thick disk component, which is emerged in this
intermediate abundance range, is somewhat over-produced. Besides this small
deviation from the observation, we conclude that the reported kinematic and
chemical properties of the Galactic halo are basically understandable in the
context of the CDM-based model for the Galaxy contraction.

\placefigure{fig-12}
\placefigure{fig-13}

\subsubsection{$V_\phi$ vs [Fe/H] relation}

Mean metal abundance of stellar populations in each region of the Galaxy
reflects at what rate star formation has proceeded, depending on the amount of
gaseous dissipation. Also, mean azimuthal velocity ($<V_{\phi}>$) of stellar
populations reflects the loss of angular momentum (in the $R$-direction)
before their birth, accompanied by gaseous dissipation.
Although these considerations are not necessarily true if the Galaxy has been
assembled from subgalactic clumps with their own formation histories, 
it might not be unreasonable to say that a relation between
[Fe/H] and $<V_{\phi}>$ for an ensemble of stars tells us about
(1) to what extent a subgalactic clump containing stars dissipates its
kinetic energy owing to gaseous cooling and dynamical friction and (2) how
the Galaxy dynamically evolves with time (e.g., Gilmore et al. 1990).
Sandage \& Fouts (1987) presented a linear relation between [Fe/H] and
$<V_{\phi}>$, in favor of the ELS scenario, whereas subsequent researchers
have disagreed with it (Norris 1986; Carney 1988; Morrison, Flynn, \&
Freeman 1990; CY; CB): the linear relation disappears at [Fe/H] $\la -1.7$.
Thus, it is intriguing to investigate what our numerical models, based on
hierarchical clustering scenario, will provide on this relation.

Figure 15 shows $<V_{\phi}>$ as a function of [Fe/H]. We have selected the
stars with their apo-galactic distances ranging from 8.5 to 17.5 kpc, to avoid
the contamination of bulge-like stars. Solid, dotted, and dashed lines denote
the stars at all heights from the plane $|Z|$, $|Z| < 1$ kpc, and
$|Z| > 1$ kpc, respectively. There are some notable features in
this figure. First, the stars with [Fe/H] $<$ $-2.2$ show a non-negative,
finite amount of rotation, where $<V_{\phi}>$ exhibits no systematic change
with [Fe/H]. This implies that the clumps having such very metal-poor stars
have not experienced large gaseous dissipation before the birth of stars;
most of such stars are born inside yet small, metal-poor clumps that are
easily disrupted by merging/interaction with other clumps
soon after their emergence.
Second, the dependence of $<V_{\phi}>$ on [Fe/H] is discontinuous at around
[Fe/H]$=-2.2$, where $<V_\phi>$ is nearly zero or somewhat negative at any
$|Z|$. As will be discussed later, we believe that this point at
[Fe/H]$\sim -2.2 $ is dynamically important in the history of the Galaxy
formation, distinguishing the nearly linear
increase of $<V_\phi>$ with [Fe/H] at [Fe/H]$>-2.2$ (as discussed below)
and apparently no systematic change of $<V_\phi>$ at less abundances.
Third,  at [Fe/H]$>-2.2$, $<V_{\phi}>$ monotonously increases with increasing
[Fe/H]. This highlights the importance of gaseous dissipation in this
$<V_{\phi}>$ vs [Fe/H] relation as follows.
The orbits of stars, initially confined inside a clump, are made circular
having high angular momentum, if the host clump suffers from gaseous
dissipation during its dynamical evolution. As the mass of the clump
increases with time due to its gravity, it suffers from further dissipation
induced by frequent merging with other clumps.
As a consequence, more metal-rich stars that are formed in the later stage
ought to have larger rotational velocity. 
Fourth, the stars at lower heights, $|Z|$, tend to have
larger $<V_{\phi}>$, implying that such stars suffer from a larger amount
of gaseous dissipation prior to their birth. 
All of these features in the $<V_{\phi}>$ vs [Fe/H] relation are basically
consistent with observations, although the inferred discontinuity in $<V_\phi>$
is located at more metal-poor abundance than the reported value
of $\sim -1.7$ dex. We here stress that these
kinematical properties of the simulated metal-poor, old stars can be affected
by long-term and somehow artificial two-body scattering resulting from
the small number of particles adopted in the simulation.

Figure 16 shows the distributions of $V_\phi$ in a given metallicity
range, for all stars (panel a) and stars at $|Z|<2$ kpc (panel b).
As is expected from the results shown in Figure 15, the very metal-poor
range with [Fe/H] $\le$ $-2$ contains stars in both prograde and retrograde
rotation; the former fraction is slightly larger.
The larger fraction of prograde stars is more easily seen at $|Z|< 2$ kpc,
which is consistent with the inferred vertical dependence of
$<V_{\phi}>$ in Figure 15. Also, the peak of the distribution
shifts toward positive $V_\phi$ for more metal-rich ranges, as is also
expected from the discussion given above.
These properties of $V_\phi$ are also qualitatively in agreement with
observations, except that the observations have revealed a finite fraction of
stars in highly retrograde rotation, at $V_\phi < -100$ km s$^{-1}$.
Absence of such stars in our results, based on simulation of an isolated
system, suggests that later accretion  of satellites plays a role in populating
them.

\placefigure{fig-14}
\placefigure{fig-15}
\placefigure{fig-16}

\subsubsection{Kinematic substructure in phase space}

If the basic physics involved in the halo formation is the merging/accretion
of subgalactic clumps, or equivalently dwarf-like satellites, and their
resultant tidal disruption, it is of great importance to search for any
signatures of fossil tidal streams in the halo, in order to
reveal the orbital paths followed by the clumps that have already merged with
the Galaxy long time ago, and thus to elucidate the merging history in the
Galactic past (Lynden-Bell \& Lynden-Bell 1995). There are a number of recent
works addressing the nature of possible candidates of globular clusters,
halo stars, and satellite galaxies with orbits tracing
the fossil tidal streams (e.g., Majewski et al. 1994; Morrison et al. 2000). 
Also, since the phase mixing of stars composing tidal debris from
the past mergers is expected to be incomplete (Helmi \& White 1999), it is
possible to find signatures of the past merging events in the form of kinematic
substructures. Recently, Helmi et al. (1999) discovered a clear
``fossil evidence'' of the past merging events in the solar neighborhood by
investigating the sample of metal-poor stars (see also CB).
They identified a statistically significant clumping of stars 
in angular momentum diagram, $L_{\rm z}$ versus  
$L_{\perp}$ = ${({L_{\rm x}}^2+{L_{\rm y}}^2)}^{1/2}$.

It is interesting to investigate whether such kinematic substructures are
also evident in the present model in which merging between subgalactic clumps
frequently occurs. Figure 17 shows the current distribution of the simulated
stars with [Fe/H] $\le$ $-1.0$ in angular momentum diagram, $L_{\rm z}$ versus
$L_{\perp}$ = ${({L_{\rm x}}^2+{L_{\rm y}}^2)}^{1/2}$. We have selected stars
at $6<R<11$ kpc and $|Z|<2.5$ kpc, to mimic the plot of nearby stars within
distance 2.5 kpc as adopted by Helmi et al. (1999) and CB.
As is evident, there exists no clear clumping in angular momentum diagram.
A possible reason for its absence is that since nearly all of merging events
occurred before $z=1$, any kinematic substructures may have already been
smeared out owing to the violently relaxing gravitational potential, as is
seen in Figure 1, 2, and 3. Also, it is possibly due to numerical effects
in the present study, i.e. unrealistically short time scale of two-body
relaxation, so that any substructures formed by merging/accretion disappear
soon after their emergence in an artificial way. 

We note that kinematic substructures are easily detectable, if some minor
merging events occur only after the formation of the entire Galaxy
structure has been completed. In such a later stage, the gravitational field
is nearly steady, so that energy exchange between clumps and background
is inefficient, compared to the stage of violent relaxation. This implies
that the reported kinematic substructure may be caused by later accretion
of a satellite, which is not taken into account in our simulations starting
from the collapse of an isolated sphere.
Thus our future large-scale simulations, which include both the kpc-scale
dynamical evolution of the Galaxy and the 100kpc-scale later accretion
processes, may well reproduce successfully the kinematical
substructure discovered by Helmi et al. (1999).

\subsubsection{Metallicity distribution}

Figure 18 shows the metallicity distributions of the stars obtained from
our simulation at $z=0$. The upper panel is for the stars selected from the
region $r>4$ kpc and $|Z|>4$ kpc, in order to isolate the halo component
as possible. Its metallicity distribution has a peak at [Fe/H]$\sim -1.6$
and a long tail at lower abundances, in good agreement with observations
(Laird et al. 1988; Ryan \& Norris 1991). Such a distribution is basically
unchanged even if we confine ourselves to the stars at larger $|Z|$.
Ryan \& Norris (1991) argued that the observed metallicity distribution
of metal-poor stars is well fitted by the prediction of the mass loss
modified model of simple chemical evolution (Hartwick 1976),
with an effective yield of $\log_{10} y_{eff}=-1.6$. Our numerical model
reveals the actual process of mass loss from the halo system, which
reduces the yield in a simple model, in the framework of the CDM theory:
gas in each of subgalactic clumps is immediately stripped after tidal
disruption of clumps, so that star formation and chemical evolution
inside clumps are ceased. As a consequence, the tidally stripped stars
have more metal-poor abundances than predicted from a simple model.

In addition to this metal-poor
component, we see a number of intermediate abundance stars ranging from
$-1.2$ to $-0.4$ dex, reminiscent of the metallicity distribution of the
Galactic thick disk (Wyse \& Gilmore 1995). This implies that the scale height
of the simulated disk component is larger than a characteristic value of
1 -- 2 kpc derived for the scale height of the thick disk (e.g., Yoshii,
Ishida, \& Stobie 1987; CY, see also Majewski 1993 for the
argument of a much larger scale height). We note that this discrepancy
is probably caused by the low spatial resolution of our simulation,
$\sim 2$ kpc.

The lower panel in Figure 18 highlights the bulge-like central component,
selected from the region $r<4$ kpc. Its metallicity distribution is dominated
by metal-rich stars, with a very long tail at lower abundances, which
agrees well with the observed distribution in the Galactic bulge
(McWilliam \& Rich 1994; Ibata \& Gilmore 1995).

\subsubsection{Age and metallicity gradients}

We plot, in Figure 19, the radial distributions of ages and metallicities
of the simulated stars at $z=0$. It follows that the metal-poor stars with
[Fe/H] $\le$ $-1.6$ show virtually no remarkable age gradient.
It is not unexpected, since most of stars with [Fe/H] $\le$ $-1.6$ are formed
at high $z$ and thus have uniformly old ages.
Also, even if an age gradient is developed in the early stage of the halo
formation, the later merging event between massive clumps smooth out the
gradient owing to violent relaxation of the gravitational field.
On the other hand, a small but finite radial gradient for metallicity
is seen ($\Delta$[Fe/H]$/\Delta r \simeq -0.012$ dex kpc$^{-1}$ for
$0 \le r  \le 25$ kpc), if {\it all} stars are concerned. The gradient
is significantly reduced if we exclude the very metal-poor stars
with [Fe/H]$<-4$.
The existence of small metallicity gradient is caused by dissipative
evolution of clumps. It is also caused by the mass-dependence of dynamical
friction: more massive and thus more metal-rich clumps can approach
more closely to the central region of the proto Galaxy, so that
more metal-rich stars are spread there after tidal disruption.

\placefigure{fig-17}
\placefigure{fig-18}
\placefigure{fig-19}

\section{Discussion}

Although there have been many discussion on advantages and disadvantages of
the ELS monolithic and SZ merger scenarios in explaining growing information
on the detailed properties of the Galactic stellar halo,
only a few studies are devoted, in the framework of the currently favored CDM
scenario of galaxy formation.
Thus, we mainly discuss about (1) success and failure of the hierarchical
clustering scenario in explaining various aspects of the observed halo, and
(2) how the scenario explains not only the stellar halo
but also the bulge and disk components in the Galaxy.

\subsection{The origin of the inner flat and outer spherical halo}

It is unlikely that either the ELS or SZ scenario alone can fully explain
the observed dual nature of the halo, showing an inner flattened part and
outer spherical part (e.g., Freeman 1996; CB). One of the possible scenarios
for the duality is to combine aspects of both the ELS and SZ scenarios,
in which the outer halo is made up from merging and/or accretion of
subgalactic objects, such as dwarf-type satellites,
whereas the inner halo is formed from dissipative, ordered contraction
on relatively short timescale (e.g., Sommer-Larsen \& Zhen 1990; 
Norris 1994; Carney et al. 1996; Sommer-Larsen et al. 1997). This hybrid
scenario is expected to explain the characteristic properties of the halo
(Carney et al. 1996; CB). If it is the case, questions then arise:
(1) what physical processes in the early
dynamical evolution of the Galaxy are actually responsible for the hybrid
formation picture and (2) how the origin of such two-fold collapse or merging
can be explained by a more sophisticated and modern theory of galaxy formation.
There have been, however, yet no theoretical studies which clearly answer these
questions.

Our numerical studies have presented the detailed aspects of the above
hybrid scenario and also other additional processes which give rise to the
dual nature of the halo (\S 3.2.2).
Namely, we have demonstrated that the outer spherical halo is formed via
essentially dissipationless merging of less massive and smaller subgalactic
clumps, whereas the inner flattened halo is formed via three different
mechanisms, i.e., dissipative merging between two massive clumps,
adiabatic compression due to later disk growth,
and rapid gaseous accretion onto the Galaxy at early epochs;
all of these three mechanisms are, for the first time, consistently
investigated in the present numerical study.

If the proposed halo formation is also the case in general disk galaxies
with different Hubble types, the present study implies the following
inter-relation between morphological type and halo structure in galaxies.
First, a later type disk galaxy (more disk-dominated galaxy) may have
a more flattened halo. As is demonstrated in our simulations, a bulge-to-disk
ratio in a disk galaxy decreases with decreasing redshift owing to later
gradual disk growth, whereas the halo structure becomes more and
more flattened because of adiabatic compression of later disk growth.
It is also remarked that in a hierarchical clustering
scenario of galaxy formation, a smaller bulge-to-disk-ratio implies that
a larger amount of gas has been accreted from the halo to disk, in
a longer time-scale (e.g., Baugh et al. 1998).
Thus, it is suggested that a later type disk galaxy, which is supposed to
accrete a larger amount of gas, may yield a more flattened halo.

Second, an earlier type disk galaxy may have a denser halo, if such a
galaxy are formed via more violent merging among more massive clumps,
as is supposed for the formation of elliptical galaxies (e.g., Barnes 1992).
Since dynamical friction is more effective for more massive clumps,
they can approach more closely to the central region of the galaxy
and then merge with one another there.
Consequently, the aftermath of the merging may build up a bigger bulge.
Since a larger number of stars are tidally stripped from these clumps
and spread over a more inner part of the halo, the aftermath may build up
a denser halo. Thus, a proto-galaxy having initially more massive
clumps may lead to a bigger bulge and denser, perhaps more metal-rich halo.
Katz (1992) has already pointed out that more substructures in a proto-galaxy
results in the formation of a denser and bigger bulge.
We thus suggest that such a mass-dependent evolution of subgalactic
clumps in the early evolution of a disk galaxy can give rise to the
correlation between bulge-to-disk ratio and halo density.

Although there are only a few observations for searching and studying
structures of stellar halos in external disk galaxies (e.g., Morrison 1999),
it is undoubtedly worthwhile to seek the predicted correlations
between galactic morphologies and halo structures, especially with
10m class large telescopes.

\subsection{Anisotropic nature of the halo velocity ellipsoid}

Internal velocity distributions of the stellar halo also provide valuable
information on the early dynamical history of the Galaxy.
The local velocity field of metal-poor halo stars in the solar neighborhood,
for which information on full space velocities are available,
is characterized by a radially elongated velocity ellipsoid
$(\sigma_R,\sigma_\phi,\sigma_Z) \simeq (141, 106, 94)$ km s$^{-1}$ (e.g., CB).
Velocity distributions in other, more remote regions are yet uncertain
because of the lack of accurate proper motion data, but the analysis of
line-of-sight velocities combined with detailed modeling of the velocity
fields allows to elucidate the global velocity distribution of the stellar
halo. Based on this strategy, Sommer-Larsen et al. (1997) analyzed the sample
of relatively bright, blue horizontal branch field stars and proposed that
halo populations show distinct changes in their velocity ellipsoid as one
moves from the inner to outer halo:
in the inner part ($r \la 10$ kpc), the radial component of the velocity
ellipsoid, ${\sigma}_{\rm r}$, is systematically larger than the tangential
component, ${\sigma}_{\rm t}$
(${\sigma}_{\rm r}$ $\sim$  140 km ${\rm s}^{-1}$ and
${\sigma}_{\rm t}$  = 90--100 km ${\rm s}^{-1}$),
whereas the outer halo at $r>20$ kpc is tangentially anisotropic
(${\sigma}_{\rm r}$ = 80--100 km ${\rm s}^{-1}$ and 
${\sigma}_{\rm t}$  = 130--150 km ${\rm s}^{-1}$).

We plot, in Figure 20, the radial dependence of ${\sigma}_{\rm r}$ and
${\sigma}_{\rm t}$ derived from the present simulation. We have selected
the stars with [Fe/H]$\le-1.0$ at $|Z|>2$ kpc to isolate the halo
component. Although the plot is not smooth due to the small number of
available stellar particles, especially in the outer part ($R>15$ kpc), there
is a signature of radially anisotropic velocity field in the inner part
($R<12$ kpc). The outer part appears to be tangentially anisotropic, but
we admit that the result is not definite because of large errors in the
calculation of velocity dispersion.

If the proposed radial dependence of ${\sigma}_{\rm r}$ and ${\sigma}_{\rm r}$
is the case, then what physical processes are responsible for such a change?
The observed radial change of velocity ellipsoid may be explained via
anisotropic, {\it dissipative} merging between protogalactic gas clouds
in a collapsing galaxy (Sommer-Larsen \& Christensen 1989; Theis 1997),
as is described below. Protogalactic gas clouds which initially have radially
eccentric orbits can plunge into the dense, inner region of the Galaxy
and collide frequently with other clouds there.
These clouds dissipate their kinetic energy and consequently never return
to the radius from which their orbital motions started. The halo stars formed
from such clouds would have rather radially eccentric orbits and
continue to orbit only within the inner part of the Galaxy.
On the other hand, protogalactic gas clouds which initially  have less
radially eccentric orbits  (or more circular orbits) cannot plunge
into the inner region and thus do not suffer from cloud-cloud collisions
so frequently. These clouds therefore preserve less eccentric
orbits and continue to locate in the outer part of the Galaxy. 
The orbits of halo stars formed from such clouds would be more circular. 
As a consequence, the inner halo which is dominated
by stars with radially eccentric orbits show more radially anisotropic
velocity dispersion, whereas the outer halo containing stars
with more circular orbits show tangentially anisotropic velocity dispersion.

An alternative picture is that the outer halo is more gradually formed from
accretion of numerous fragments, such as satellite galaxies,
having less eccentric orbits. To be more specific, most of stars
now observed in the outer part of the halo are confined inside satellites,
which are initially located at large distance from the Galaxy, having large
orbital angular momentum, and the tidal disruption of such satellites occur
at relatively later epoch, as in the case of the Sgr dwarf galaxy (Ibata,
Gilmore, \& Irwin 1984). Since the orbits of the stars somehow preserve
less eccentric orbits of their progenitor satellites,
the outer halo is made tangentially anisotropic.

We note that in the present simulation, the total number of subgalactic
clumps are rather small owing to the small number of the adopted initial
gaseous particles. Accordingly, {\it anisotropic} dissipative merging
due to orbit-dependent collisions between clumps is
suppressed, especially in the outer region of the system.
Also, we have investigated an initially {\it isolated} system, without
taking into account the effect of later accretion of satellites;
merging and accretion events cease at relatively early epoch.
Thus, it is yet unsettled from the present simulation, which picture
described above is likely to explain the observed radial change of
velocity dispersions. We are planning to revisit this issue, based on
high-resolution simulations covering both 10 kpc-scale dynamical
evolution of the Galaxy and 1 Mpc-scale merging history of
dwarf-type galaxies in the forming Local Group.

\subsection{The Galactic turning point}

Many researchers have attempted to settle an issue as to how the rotational
velocities of field halo stars, $V_\phi$, change with their metallicities,
since ELS's work, showing larger $V_\phi$ for more metal-rich stars
(e.g., Gilmore et al. 1990).
Sandage and Fouts (1987) pointed out that there is a clear continuation
of a {\it smooth} relation between $<V_\phi>$ and [Fe/H] and suggested
that this correlation is an evidence supporting the monolithic collapse
picture by ELS. On the other hand, using 1200 stars with [Fe/H] $\le$ $-0.6$, 
Norris (1986) argued that a continuously smooth change of $<V_\phi>$ is
not the case over an entire range of [Fe/H]: he found a sharp discontinuity 
at [Fe/H]$\sim -1.4$ below which there is no explicit dependence of $<V_\phi>$
on [Fe/H] (see also  Norris \& Ryan 1989; Beers \& Sommer-Larsen 1995; CY; CB).
He concluded that this result is consistent with the SZ picture that
numerous protogalactic fragments merge with one another to form the Galaxy.
Recent more detailed analysis by CB, using 1203 stars with revised information
on their kinematics and metallicities, has confirmed a clear discontinuity
in the relation between $<V_\phi>$ and [Fe/H] at [Fe/H] $\sim$ $-1.7$. 
CB also found, for stars with $-2.4\le$[Fe/H]$\le-1.9$, a systematic decrease
of $<V_\phi>$ with increasing distance from the Galactic plane, $|Z|$.
Based on these results, they concluded that (1) the rapid collapse picture
by ELS is inconsistent with the discontinuous relation between $<V_\phi>$ and
[Fe/H] and (2) totally chaotic merging in the SZ picture also has
difficulties in explaining the vertical gradient of $<V_\phi>$.

The present numerical simulations based on hierarchical clustering scenario
have successfully reproduced {\it qualitatively}
both the discontinuous change of $<V_\phi>$ with [Fe/H] and the vertical
gradient of $<V_\phi>$:
dissipative merging of subgalactic clumps in the early epoch of galaxy
formation ([Fe/H] $\le$ $-1.7$) and the subsequent gradual accretion of gas
([Fe/H] $>$ $-1.7$) in the later stage give a plausible explanation
for the observed properties of $<V_\phi>$. 
The problem remained in our model is that the present simulation indicates the
discontinuity at [Fe/H] $\sim$ $-2.2$, which is systematically more metal-poor
than the observed value of $\sim$ $-1.7$.
Since we regard the point at [Fe/H] $\sim$ $-1.7$ as a crucial epoch
for understanding the Galaxy formation, as is described below,
we will attempt to solve this discrepancy in our future work, by elaborating
numerical models and investigating a full coverage of model parameters
(Bekki \& Chiba 2000b).

CB also showed that there is a concentration of high-$e$ stars at
[Fe/H] $\sim$ $-1.7$, which gives rise to a nearly zero mean rotation
of the stars at this metallicity. Except at this discontinuous point,
$<V_\phi>$ stays positive, 30--50 km s$^{-1}$, at [Fe/H] $<$ $-1.7$,
whereas at [Fe/H] $>$ $-1.7$, $<V_\phi>$ linearly increases with increasing
[Fe/H]. As a consequence, there appears a hollow-like feature at
[Fe/H] $\sim$ $-1.7$ in the $<V_\phi>$ vs [Fe/H] relation
(see Figure 3 in CB). 
It remains uncertain how this hollow-like feature, consisting of high-$e$
stars, is created. In conjunction with this,
it is equally important to explain the properties of $<V_\phi>$ below and
above [Fe/H] $\sim$ $-1.7$, to elucidate physical processes involved in
the formation of the halo and (thick) disk components, respectively.
Thus, we believe that this hollow-like feature is a key to understanding
how the stage of halo formation is discontinuously changed to that of
disk formation.

Here we attempt to explain the origin of this discontinuous change of
$<V_\phi>$ and more details will be studied in Bekki \& Chiba (2000b).
Our models indicate that just at the epoch when the Galaxy as a whole has
the mean metallicity of [Fe/H] $\sim$ $-1.7$ (or the epoch when the Galaxy
is dynamically dominated by massive clumps with [Fe/H] $=$ $-1.7$),
the last merging event between two massive clumps takes place.
This merging event is totally violent and dissipative, as the mass ratio of
these clumps is close to 1. Many of the stars after disruption preserve
orbital angular momentum of clumps, thereby build up an inner halo with
somewhat prograde rotation. On the other hand, strong gaseous compression
at the moment of merging give rise to newly born stars with
[Fe/H]$\sim -1.7$. Such stars may have less amount of angular momentum
because they are born within the strongly compressed region between
both clumps. 
Furthermore, the orbits of such stars are randomized by violent relaxation
in the final phase of merging of. Consequently, the orbital eccentricity of
each of the stars becomes rather large.
Thus, in this manner, we may be able to reproduce the observed
discontinuous of $<V_\phi>$ at [Fe/H] $\sim$ $-1.7$: it is closely associated
with the gaseous compression and violent relaxation due to the last merging
event between subgalactic clumps. Since the basic
process involved in the evolution of the Galaxy appears to be different
between before and after this point, we would like to call it as
{\it the Galactic turning point} in the Galaxy history.

The above picture provides the following further implications for the Galaxy
formation (Bekki \& Chiba 2000b).
First, the Galactic bulge may have been formed at this epoch, owing to 
strong starburst associated with the last merging event.
A large amount of compression and dissipation of gas at this point
may induce an efficient radial inflow of gas and subsequently trigger
strong starburst. The stars born by this process, which are well enriched
because of rapid chemical evolution during starburst, may form a large
fraction of the bulge. Some fraction of metal-poor, old stars, which are
initially inside the clumps, are also expected to comprise the bulge.
Therefore, if the bulge is formed in this manner, it is basically an
old system having both metal-poor and metal-rich stars.
Second, the formation of the halo component at the turning epoch may play a key
role in the formation of the first {\it thin} disk, which is {\it not} the
present-day thin disk as will be described later. This is inferred from
the following. After the most part of the halo is formed, the gravitational
potential stays nearly fixed, under which low density gas stripped during
early merging events among numerous small clumps are fallen onto the
equatorial plane, because there is no centrifugal barrier in this direction.
This gradual accretion of gas may give rise to a thin disk in the equatorial
plane.

\subsection{Minor merging and thick disk formation}

There are variously different scenarios for the formation of the
thick disk component with a scale height of about 1 kpc
(e.g., Gilmore et al. 1990; Majewski 1993). These include,
(1) a rapid increase in the dissipation and star formation rate because
of more rapid cooling of enriched gas with [Fe/H] $>$ $-1$
(Wyse \& Gilmore 1988),
(2) violent dynamical heating of an early thin disk induced by
minor satellite merging at high redshift
(e.g., Quinn et al. 1993),
(3) long-term kinematic diffusion of stars formed initially
in a thin disk, and (4) direct satellite accretion of thick disk material 
(Statler 1989).
We here focus on the second minor merging scenario, because minor merging
is expected to occur frequently in the hierarchical clustering scenario.
In this scenario, the orbits of subgalactic clumps or satellites are
circularized and are gradually fallen into the center of the disk, owing to
dynamical friction and tidal force from the disk.
The disk stars are heated dynamically by minor merging, and the aftermath
is similar to the observed spatial structure and kinematics of the Galactic
thick disk. The issues addressed here are (1) when such minor merging
occurs in order to explain both the metallicity distribution 
and total mass of the thick disk, and
(2) if the merger scenario explains the observed (small) metallicity gradient
of the thick disk.

We here propose  that if the minor merging between
a first thin disk and a satellite galaxy (or subgalactic clump)
occurs at [Fe/H] $\sim$ $-0.6$, this scenario can settle the above two issues.
After the bulge formation via the last merging event between massive
clumps, the first thin disk begins to form and the chemical evolution proceeds,
starting from an initial abundance of [Fe/H] $\sim$ $-1.7$.
If so, the stars formed inside the disk have metallicities larger than $-1.7$.
If the minor merging related to the thick disk formation
occurs when the mean metallicity of stars reaches around $-0.6$,
the thick disk developed from dynamical heating is expected to contain stars 
with the metallicity ranging from $-1.7$ to $-0.6$. The total mass of the
thick disk may be determined by the total gas mass accumulated during the
formation of the first thin disk.
Furthermore, since both the first disk and the subgalactic clump
have a large amount of gas, this minor merging occurs basically in a
dissipative manner, where any spatial gradient in the metallicity of
the thin disk remains preserved, in contrast to totally dissipationless merging
(e.g., Bekki \& Shioya 1998). 
Therefore, the vertical metallicity gradient in the first thin disk
may be preserved in the thick disk.

Thus, this minor merger scenario may explain, without obvious difficulties,
the observed properties of the Galactic thick disk.
It is noted that starburst is triggered by
 minor merging (e.g., Mihos \& Hernquist 1994).
Furthermore, recent observational results on the relation between
the vertical velocity dispersion of disk stars, $\sigma_z$, and their ages
have demonstrated that there is a sudden increase of $\sigma_z$
for disk stars with ages of 9 -- 12 Gyr (Quillen \& Garnett 2000).
If such disk populations have the metallicity of $\sim$ $-0.6$
and also show some evidence of the past starburst at [Fe/H] $=$ $-0.6$,
the validity of the minor scenario is greatly strengthened.
In this regard, it is of interest to seek the more precise determination of
the age-metallicity distribution of disk stars with the ages ranging from
9 to 12 Gyr and the mean metallicity of $\sim$ $-0.6$, to assess the
minor merger scenario for the formation of the thick disk.

\subsection{A rough sketch of the Galaxy formation}

Based on the present numerical studies, we here summarize a possible scenario
of the Galaxy formation in the framework of hierarchical merging of
subgalactic clumps. To describe each evolutionary process step by step,
we use the mean metallicity of gas inside the entire Galactic space as a
rough indicator of clock.

\subsubsection{[Fe/H] $<$ $-1.7$}

Just after the formation of numerous small clumps owing to the non-linear
growth of initial small-scale density perturbations, these clumps tidally
interact and merger with one another. Before the mean metallicity of gas
reaches $-1.7$ dex, the stars inside these small clumps are tidally stripped
during interaction and merging, and are subsequently spread over the outer
part of the halo. This outer halo, consisting of old and metal-poor stars,
has a spherical density distribution with no systematic rotation.
Along with this
process, two more massive clumps, having nearly equal masses, have been
developed from multiple merging of small clumps, and chemical evolution
proceeds inside them.

This picture for the formation of the outer halo is basically the same as
the SZ scenario. The time scale involved in this process
is more than 1 Gyr, which is slow compared with a typical free-fall time
of 10$^8$ yr in the Galaxy, for the reason that gravitational collapse is
delayed owing to the expansion of the background Universe.

\subsubsection{[Fe/H] $\sim$ $-1.7$}

When the gaseous metallicity inside the two massive clumps reaches
[Fe/H] $\sim$ $-1.7$, they merge with each other and the aftermath after
their disruption forms an inner part of the halo.
During this merging event, strong gaseous compression and violent relaxation
take place,
leaving newly born stars with less amount of angular momentum.
As a result, such numerous stars having metallicity [Fe/H] $\sim$ $-1.7$
show high-$e$ orbits and thus small mean rotation.
Also, rapid radial infall of gas to the center of the system is induced
by efficient dissipation, triggering starburst in a bulge-like component. 
The stars born by this starburst activity are well enriched, as the
chemical evolution rapidly proceeds in this central component.
If some globular clusters are associated with the clumps, they are
also stripped from the clumps and may form the system of halo globular
clusters having metal-poor stellar populations.
These clusters may also have high $e$ and small $<V_\phi>$.

\subsubsection{$-1.7$ $<$ [Fe/H] $<$ $-0.6$}

Successive accretion of the surviving, small clumps onto the equatorial
plane of the Galaxy continues, even after the end of the last merging
event. Gas in these clumps is accumulated in the plane, thereby making up
a thin disk. In addition, some more gas associated with HVCs may also accrete
from outside and increase the mass of the disk.
Star formation and chemical evolution give rise to enriched disk stars having
large rotational velocity. As gaseous dissipation proceeds and thus drives
a gradual accretion in the radial direction, more metal-rich stars
are expected to rotate more rapidly, leading to the linear increase of
$<V_\phi>$ with [Fe/H].
We estimate the total mass of this first thin disk as $\sim 10$ \% of the
mass of the present-day stellar system in the Galaxy
($\sim$ 6.0 $\times$ $10^9$ $M_{\odot}$).

\subsubsection{[Fe/H] $\sim$ $-0.6$}

When the mean metallicity of the thin disk reaches [Fe/H] $\sim$ $-0.6$,
one or some relatively massive satellites, having the mass of a few \% of
the present Galaxy and with high angular momentum, may slowly spiral toward
the Galaxy. Dynamical friction and tidal force from the thin disk make
the orbit of the satellite become more circular and approach to the disk plane.
Then, the minor merging takes place, which results in the dynamical heating
of the disk, preferentially in the $Z$ direction. This minor merging
thus develops the disk having a large scale height, perhaps comparable to
the present-day thick disk. This disk component contains the stars with 
$-1.7$ $\la$ [Fe/H] $\la$ $-0.6$ originally included in the first thin disk,
and also some more metal-rich stars born in the process of gaseous dissipation
during merging. Gaseous dissipation also induces radial flow of gas to the
central region and triggers nuclear starburst, so that both the mass and
metallicity of the bulge component are increased.

If the sinking satellite contains globular clusters, they are tidally
stripped, and form the system of clusters near the plane, which may
correspond to the observed disk clusters in the Galaxy.
Since the orbital angular momentum of the satellite is transferred to
the stripped globular clusters, the cluster system developed after this minor
merging event shows a net rotation, in contrast to the halo clusters.

\subsubsection{[Fe/H] $>$ $-0.6$ }

Along with or after the formation of the thick disk, the replenishment of gas
near the disk plane may continue, in the form of cooling flow from rarefied
gas in the halo or HVCs from outside, thereby making up another thin disk
component. Since this accretion process progresses in a quiescent manner,
the disk remains thin. As the stars in this thin disk have been formed from
a continuously replenished gas, their metallicity distribution may be exempted
from the so-called G dwarf problem. This disk component may correspond to
the present-day thin disk with a scale height of $\sim 350$ pc.

\subsection{Future work: Extragalactic halo astronomy}

It is intriguing to examine whether our models aimed at understanding the
Galactic stellar halo are also applied to external disk galaxies. For instance,
the following questions are immediately concerned: (1) Are there any difference
in dynamical and chemical properties of stellar halo components among
disk galaxies with different Hubble types?  (2) Are there any physical
correlation between, for example, bulge size and mean halo density?
(3) How can merging processes of subgalactic clumps control the final
dynamical and chemical properties of stellar halos in other galaxies?
We believe that extensive observational studies of external stellar halos
to answer these questions will shed new insight into the formation of
general disk galaxies, in addition to further information on the Galaxy
as provided by next generation astrometric satellites
(e.g., {\it FAME} and {\it GAIA}).
Also, further modeling of a halo component over a full range of model
parameters (Bekki \& Chiba 2000b) is important to understand how a variety
of halos are developed in other disk galaxies.

Here, we point out the following two aspects to be investigated thoroughly,
as a possible route to the general picture for the halo formation.

One is to extract the detailed projected light distributions of stellar halos
in external disk galaxies. In the case of M31's stellar halo, its surface
luminosity distribution follows the so-called $R^{1/4}$ law, where $R$ is the
projected distance from the center (e.g., Freeman 1999). This is in contrast
to the surface density distribution expected from the power-law volume density
of $\rho (r) \sim r^{-3.5}$ in the Galaxy. Furthermore, it is suggested that
the mean density of M31's stellar halo is ten times higher than that of the
Galactic halo (Guhathakurta, Reitzel, \& Grebel 2000).
These observed difference between the Galactic and M31's halos may reflect
the different formation history of halos in these galaxies. In addition,
Morrison (1999) suggested a marginal evidence that there is a variety
of structures in the stellar halos of external edge-on disk galaxies with
different Hubble types. In this regard, we note that our preliminary
calculations, as will be reported in our next paper, suggest that final
structures of stellar halos depend sensitively on initial conditions of
protogalaxies, such as spin parameters and amplitudes of CDM perturbations
(Bekki \& Chiba 2000b).
Thus, it is of great interest to conduct further and more detailed observations
of stellar halos in many disk galaxies and compare with such theoretical
results, in order to clarify the formation process of the stellar halo
as well as the origin of the Hubble sequence.
Such studies should be eagerly pursued with 10m class telescopes, such as
{\it SUBARU} and {\it VLT}, taking advantage of their ability to detect
faint images. We believe that extracting global distributions of stellar
halos and also some spatial substructures using large telescopes will
shed new insight into our understanding of galaxy formation.

The other is to determine the color distributions of stellar halos with
different Hubble types, in addition to the light distributions.
These information will provide us with both the initial mass function (IMF)
in the formation of the stellar halo and the nature of halo populations
as indicative of their age and metallicity distributions.
Recent observational studies on the photometric properties of NGC 5907
have suggested that peculiar colors in its halo component imply
the different shape of IMF as well as the different lower and upper mass
limits in IMF, from the widely accepted Salpeter IMF (e.g., Zepf et al. 2000).
It is thus important to conduct further observations of color distributions,
to elucidate any differences in the inferred IMF among disk galaxies with
different Hubble types and any evidence of the relation between the shape
of IMF and the structural parameters of disks such as bulge-to-disk ratio.
It is similarly important to elaborate chemodynamical models for the
formation of stellar halos, taking into account evolutionary population
synthesis. The detailed comparison between such models and observations
will allow us to clarify any relations between early dynamical processes
and photometric properties of halos, so that we will be able to obtain a more
advanced picture on the origin of the Hubble sequence.

\subsection{Limitations of the present model}

In the present study, we have adopted an isolated rotating sphere superposed by
small-scale density fluctuations for the investigation of the halo formation.
Owing to this rather specific initial condition of the proto-Galaxy,
both effects of satellite (or gaseous) accretion onto the Galaxy and major
merging at lower redshifts are suppressed in the present simulations.
Since these physical processes may affect the formation histories of the
stellar halo, it is worthwhile to consider how these processes
affect our results. We discuss below three possible aspects involved in
later accretion or merging events onto the simulated isolated system, and
firmer predictions await planned, more extended simulations.

First, later accretion of dwarf-type satellites can change the derived halo
structure, in particular the slope of the density profile.
Stellar components of satellites, which are tidally
disrupted by the Galaxy at low redshifts, can be dispersed into
the (mainly) outer part of the Galaxy and thus add up stars there.
The amount of the change in the density profile of the halo depends on
the detailed accretion process of satellites, e.g., how closely satellites
can approach the inner part of the Galaxy before their tidal disruption.
Second, such later accretion processes can form a statistically significant
clumping of stars in angular momentum diagram, $L_{\rm z}$ versus
$L_{\perp}$ = ${({L_{\rm x}}^2+{L_{\rm y}}^2)}^{1/2}$,
as was discovered by Helmi et al. (1999) in the solar neighborhood.
Third, the time evolution of the bulge-to-disk-ratio discussed here can be
modified if the effect of later accretion or merging onto the Galaxy is taken
into account. Such events can rapidly increase the bulge-to-disk-ratio
at low redshifts, because very efficient radial transfer of
gas resulted from merging can trigger the nuclear starbursts that eventually
form young bulge components, and because sinking massive satellites which reach
the central region of the Galaxy can add up stars in the bulge.

\section{Conclusions}

We have performed numerical simulations for the formation of the Galactic
stellar halo, based on the currently favored CDM theory of galaxy formation.
The main results are summarized as follows.

(1) Basic physical processes involved in the formation of the stellar halo
are described by both dissipative and dissipationless merging of subgalactic
clumps and their tidal disruption in the course
of gravitational contraction of the Galaxy. A large fraction of
metal-poor stars ($\sim$ 50 \%), making up the present-day halo,
are already formed at $z$ $>$ 2.6, inside the clumps which are developed
from small-scale CDM density perturbations,
and are then tidally stripped from the clumps by successive merging and
spread over the halo at 1.5 $<$ $z$ $<$ 3.0.
Although these halo populations are generally old, the formation of
the entire halo structure is relatively recent, at $z$ $\sim$ 1.5.

(2) The radial density profile of the simulated halo is similar to 
the observed power-law form of $\rho (r)$ $\sim$ $r^{-3.5}$,
implying that hierarchical merging of clumps is also responsible
for the characteristic radial distribution of halo stars.
We also see the lack of the simulated stars along the polar axis,
which reflects the anisotropic effect of dynamical friction on the orbital
motions of clumps.

(3) Our numerical model successfully reproduces the observed metallicity
distributions of the halo and also bulge components.
The simulated stars with [Fe/H] $\le$ $-1.6$ show virtually no radial
gradient for stellar ages, because the intrinsic age spread among such stars
is small and also the violent relaxation at the epoch of the last merging
event between two massive clumps smoothes out any gradients.
A small radial gradient is seen for stellar metallicities, in particular,
in the inner region of the halo. This may reflect the dissipative process
of gas during the merging among clumps.

(4) The inner flattened and outer spherical halo is successfully
reproduced qualitatively by the present model, based
on the hierarchical clustering scenario.
The outer spherical halo is formed via essentially dissipationless
merging of less massive and smaller subgalactic clumps
whereas the inner flattened one is formed via three different mechanisms,
i.e., dissipative merging between more massive clumps,
adiabatic compression due to  later disk growth,
and gaseous accretion onto the equatorial plane.

(5) For the simulated stars with [Fe/H] $\le$ $-1.0$,
there is no strong correlation between metal abundances and orbital
eccentricities, in good agreement with the recent observations.
Moreover, the observed fraction of the low-eccentric stars
is reproduced correctly for [Fe/H] $\le$ $-1.6$ and approximately
for the intermediate abundance range of $-1.6$ $<$ [Fe/H] $\le$ $-1.0$.

(6) The simulated stars with [Fe/H] $\le$ $-2.2$ show a small systematic
rotation without explicit dependence on [Fe/H], whereas the stars with
higher metallicities show the linear increase of $<V_\phi>$ with increasing
[Fe/H]. These properties of $<V_\phi>$ are generally in agreement with
observations, although the observed $<V_\phi>$ turns to increase with
[Fe/H] at somewhat higher [Fe/H]. Also, as is deduced from the observation,
the simulated stars at lower heights from the plane, $|Z|$, show systematically
larger $<V_\phi>$.

(7) We have not found any clear clumping of simulated metal-poor stars
in angular momentum space at $z$ $=$ 0,
though merging of subgalactic clumps plays a decisive role in the formation
of the halo. A possible reason for its absence may be due to either
numerical effects such as unrealistically short time scale of
two-body relaxation or our specific initial condition of an isolated
protogalaxy, which neglects later accretion of satellite galaxies.

Based on these results, we have discussed how early processes of both
dissipationless and dissipative merging of subgalactic clumps can reproduce,
plausibly and consistently, the recent observational results on the
Galactic halo. Also, we have presented a possible scenario for the formation of
the entire Galaxy structure, including the stellar halo, bulge, and disk
components.
It has also been remarked that further observational studies of faint halo
components in external disk galaxies, using 10m class large telescopes,
will provide us with invaluable information on the origin of the Hubble
sequence in the context of hierarchical clustering scenario of
galaxy formation.

\acknowledgments
We are  grateful to the anonymous referee for valuable comments,
which contribute to improve the present paper.
K.B. acknowledges the Large Australian Research Council (ARC).
We are grateful to  Edmund Bertschinger for allowing us to
use the COSMICS (Cosmological Initial Conditions and
Microwave Anisotropy Codes), which is a package
of fortran programs for generating Gaussian random initial
conditions for nonlinear structure formation simulations.

\clearpage


\begin{figure}
\figcaption{
Distribution of dark matter particles of the forming Galaxy projected onto
the $x$-$z$ plane, at each of redshifts indicated in the upper left corner
of each panel. Each frame measures 176 kpc on a side.
\label{fig-1}}
\end{figure}

\begin{figure}
\figcaption{
The same as Figure 1 but for gaseous particles.
\label{fig-2}}
\end{figure}

\begin{figure}
\figcaption{
The same as Figure 1 but for stellar particles formed from gas.
\label{fig-3}}
\end{figure}

\begin{figure}
\epsscale{1.0}
\plotone{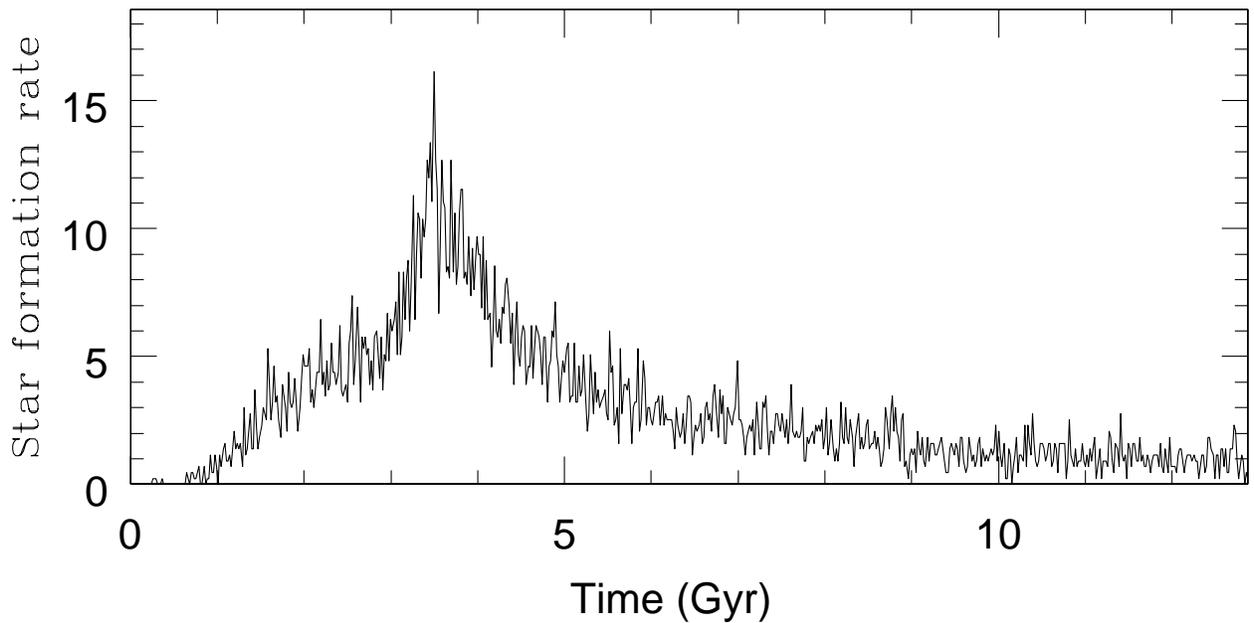}
\figcaption{
The star formation history of the present model
in units of $M_{\odot}$ ${\rm yr}^{-1}$.
\label{fig-4}}
\end{figure}

\begin{figure}
\epsscale{1.0}
\plotone{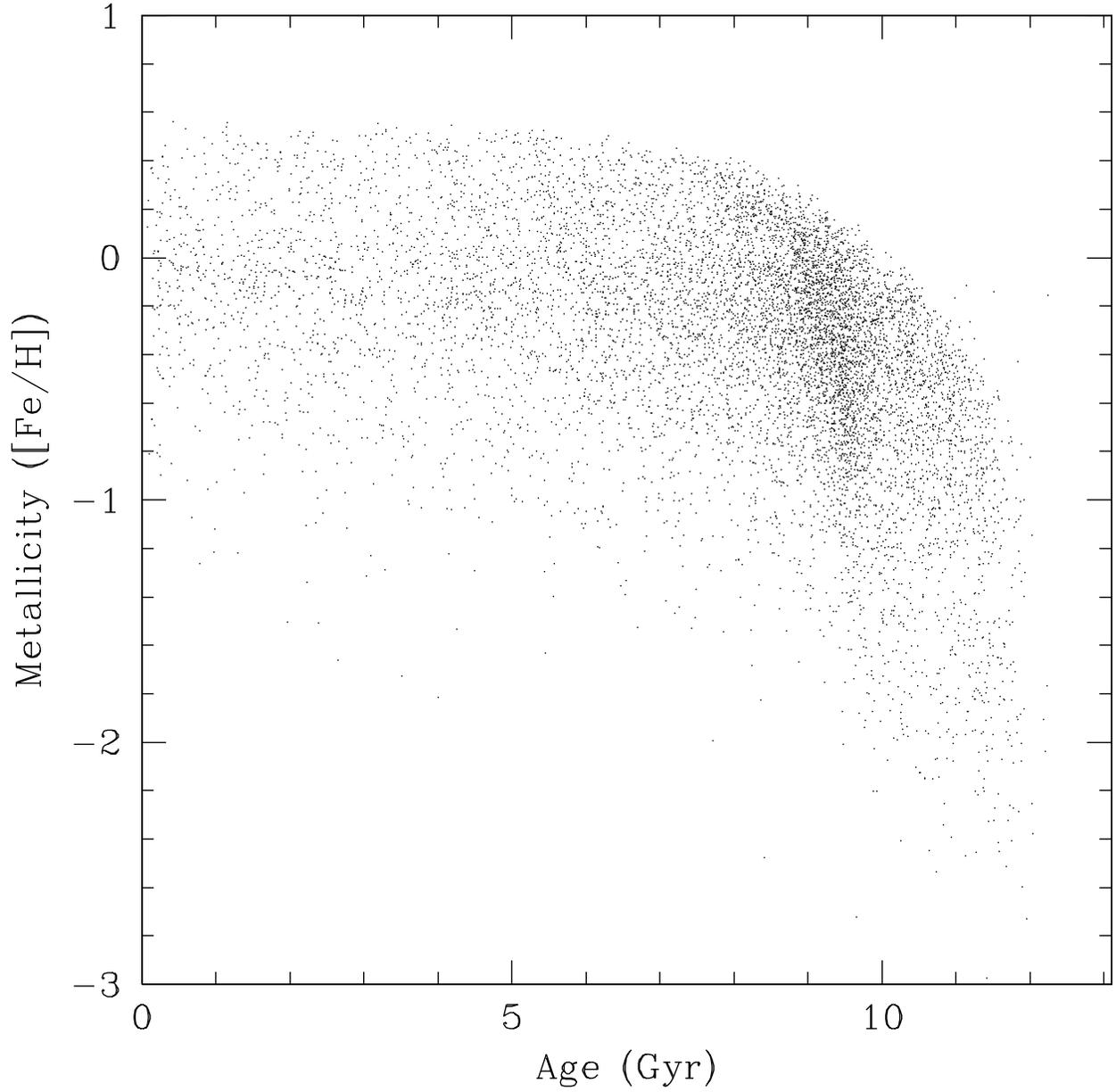}
\figcaption{
Age-metallicity diagram of the stars in our simulation.
Note that there is a rather large dispersion in metallicities
for a given age.
\label{fig-5}}
\end{figure}

\begin{figure}
\epsscale{1.0}
\plotone{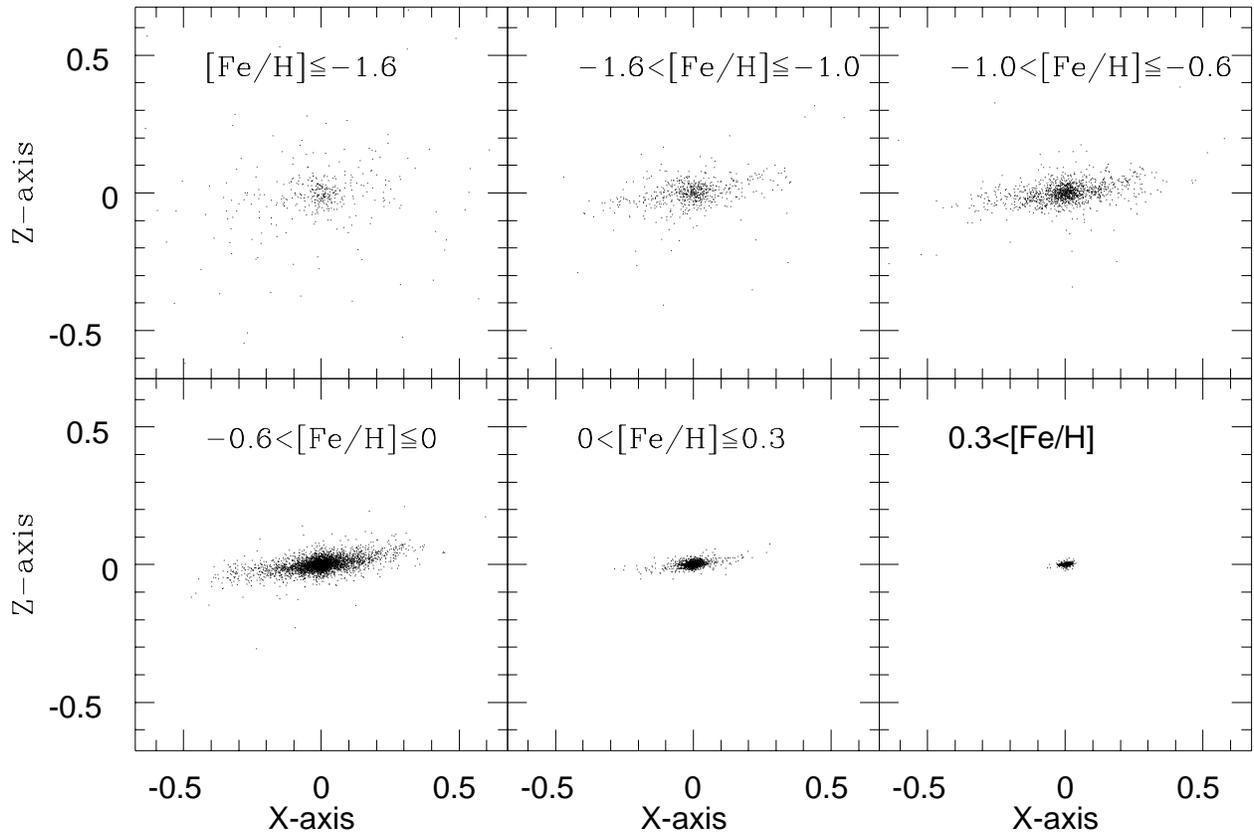}
\figcaption{
Final distribution of the stars for a given metallicity range,
projected onto the $x$-$z$ plane at $z$ $=$ 0 ($T$ = 13 Gyr).
The metallicity range is indicated in each panel.
Each frame measures 58.5 kpc on a side.
\label{fig-6}}
\end{figure}

\begin{figure}
\epsscale{1.0}
\plotone{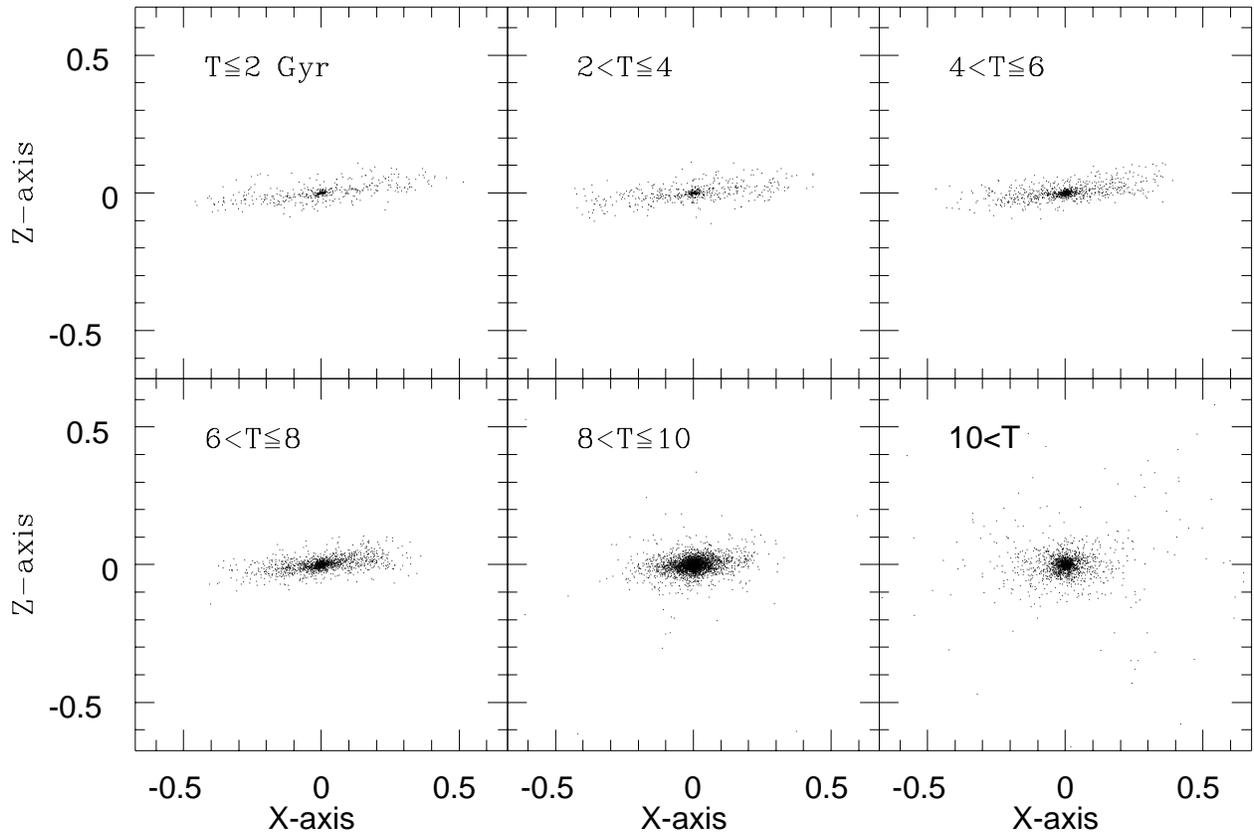}
\figcaption{
Final distribution of the stars for a given age range,
projected onto the $x$-$z$ plane at $z$ = 0 ($T$ = 13 Gyr).
The age range is indicated in each panel.
Each frame measures 58.5 kpc on a side.
\label{fig-7}}
\end{figure}

\begin{figure}
\epsscale{1.0}
\plotone{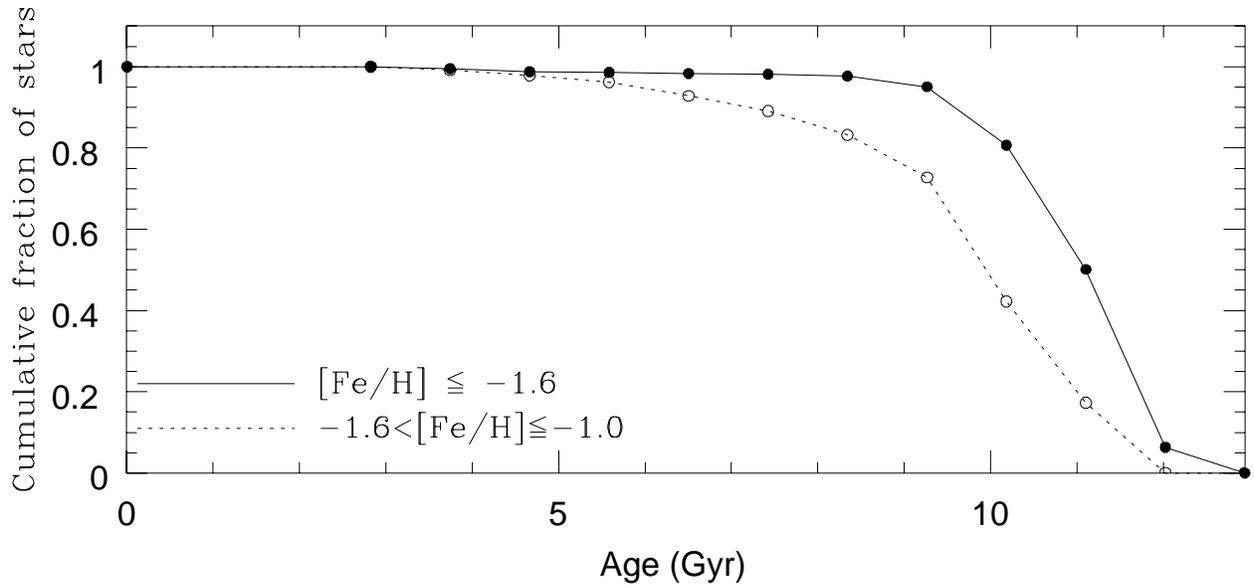}
\figcaption{
Number fraction of the stars having older ages than a given age (or cumulative
age distribution), for the metallicity ranges of
[Fe/H] $\le$ $-1.6$ (solid line) and $-1.6$ $<$ [Fe/H] $\le$ $-1.0$ (dotted
line), respectively.
This figure also denotes the number fraction of the stars which have already
been formed by the given epoch. 
\label{fig-8}}
\end{figure}

\begin{figure}
\epsscale{1.0}
\plotone{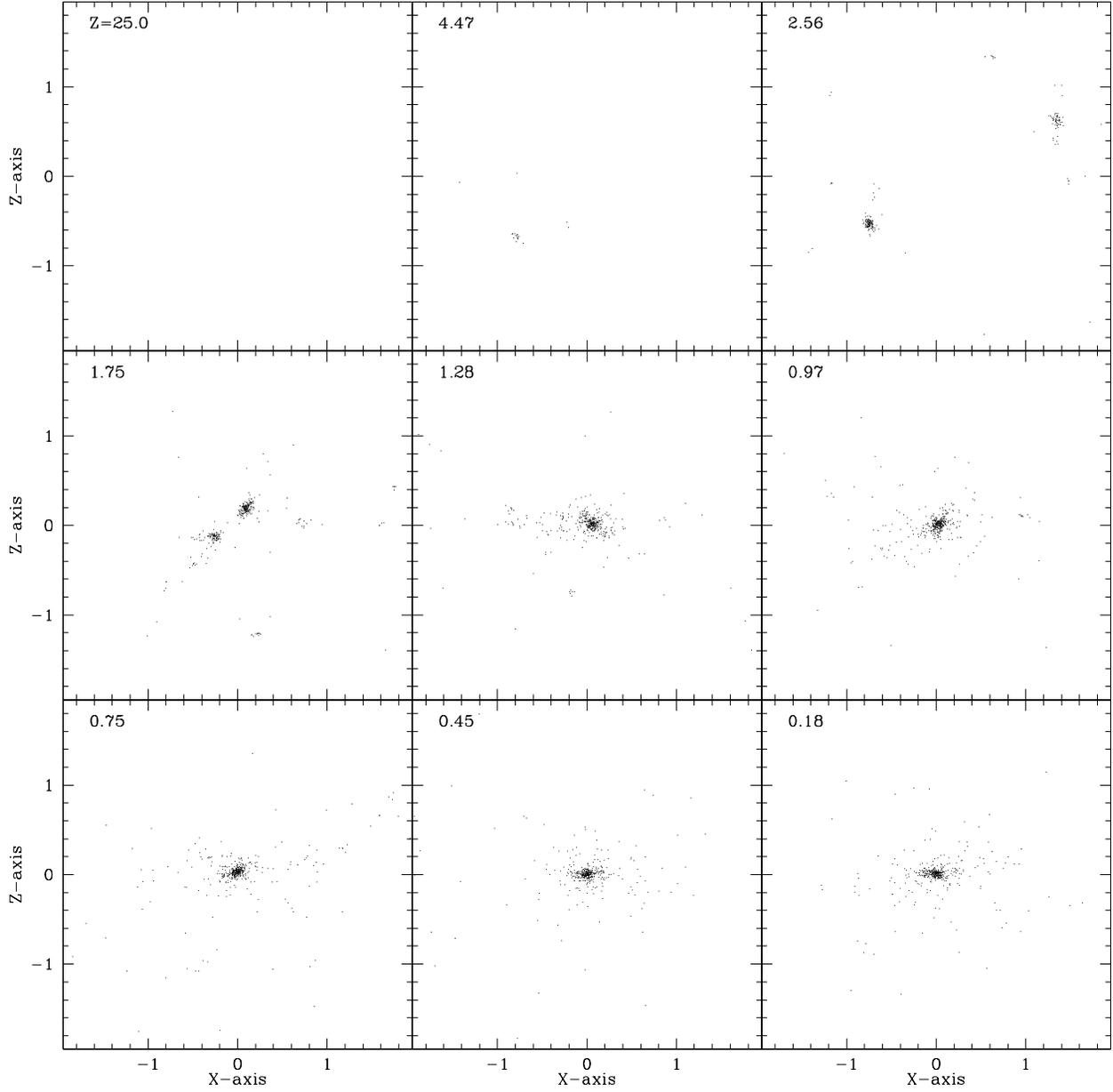}
\figcaption{
Spatial distribution projected onto the $x$-$z$ plane at each redshift,
for stellar particles that finally become metal-poor halo
component with metallicity [Fe/H]$\le-1.6$ at $z=0$.
Each frame measures 176 kpc on a side.
\label{fig-9}}
\end{figure}

\begin{figure}
\epsscale{1.0}
\plotone{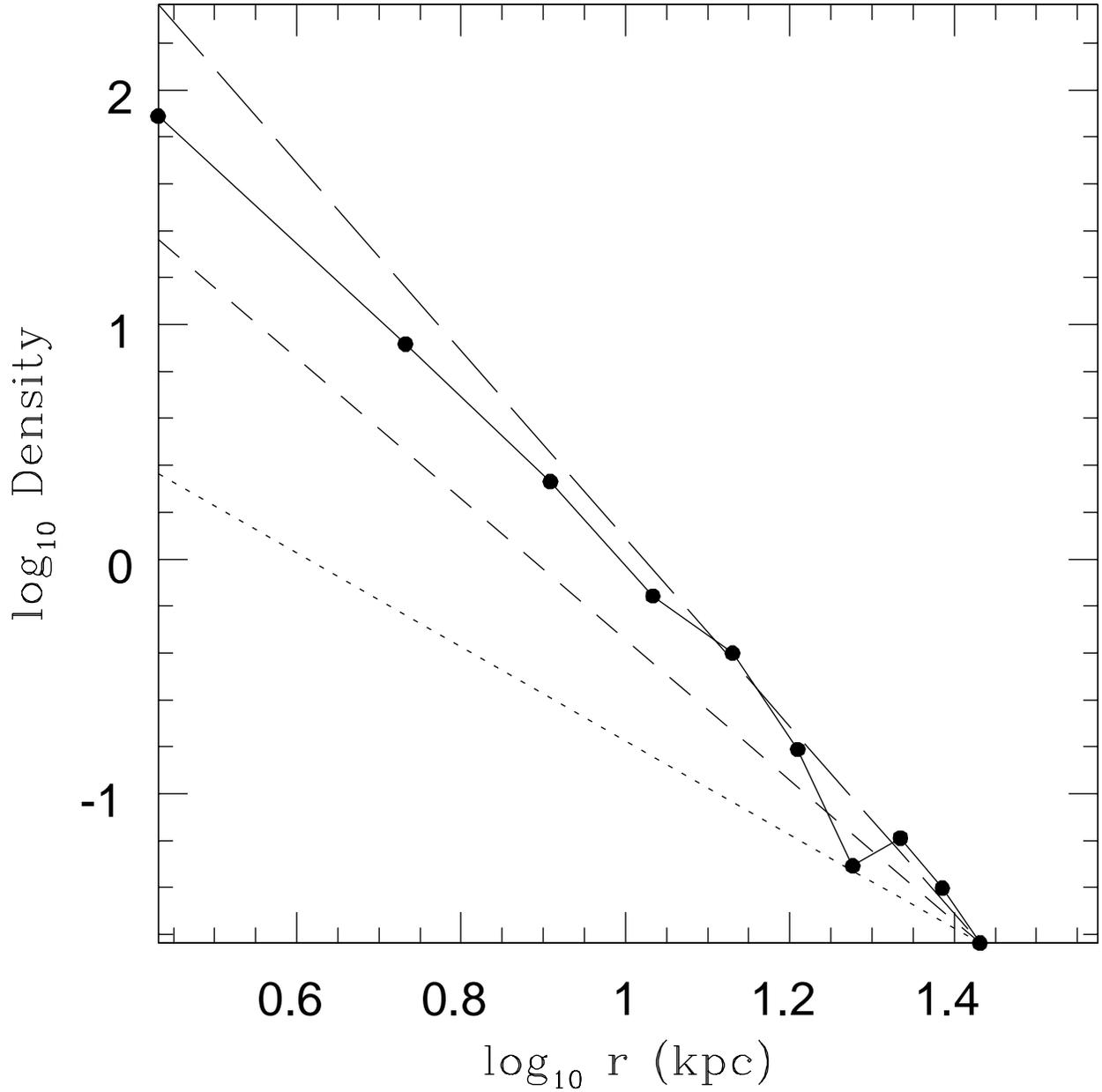}
\figcaption{
Density profile of the simulated stars with [Fe/H] $\le$ $-1.6$
as a function of distance from the center, $r$
(solid line with filled circles).
For comparison, the density profiles, $\rho (r)$ $\sim$ $r^{-2}$,
$r^{-3}$, and $r^{-4}$, are superimposed by dotted, short-dashed,
and long-dashed lines, respectively.
\label{fig-10}}
\end{figure}

\begin{figure}
\epsscale{0.8}
\plotone{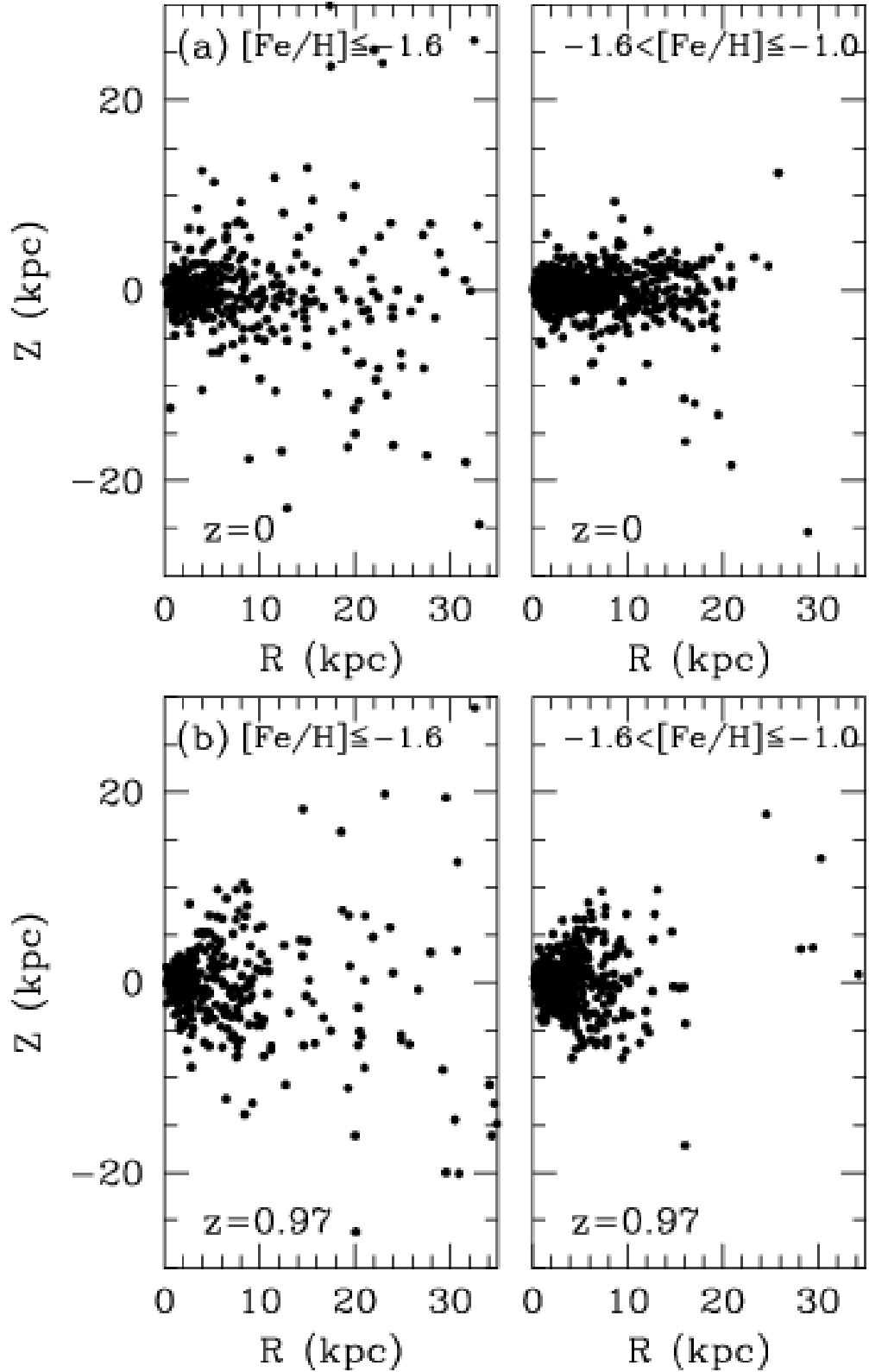}
\figcaption{
Meridional distribution of the stars (within $R = 35$ kpc and $|Z| = 35$ kpc)
at $z$ $=$ 0 (a) and $z$ $=$ 0.97 (b). Left and right panels show the stars
with [Fe/H] $\le$ $-1.6$ and $-1.6$ $<$ [Fe/H] $\le$ $-1$, respectively.
The redshift $z$ $=$ 0.97 corresponds to the epoch after $\sim$ 1 Gyr
passage of the merging event between two massive subgalactic clumps.
Note that the halo appears more flattened at $z$ $=$ 0 than at $z$ = 0.97,
as is clearly seen for $-1.6$ $<$ [Fe/H] $\le$ $-1$.
\label{fig-11}}
\end{figure}

\begin{figure}
\epsscale{0.8}
\plotone{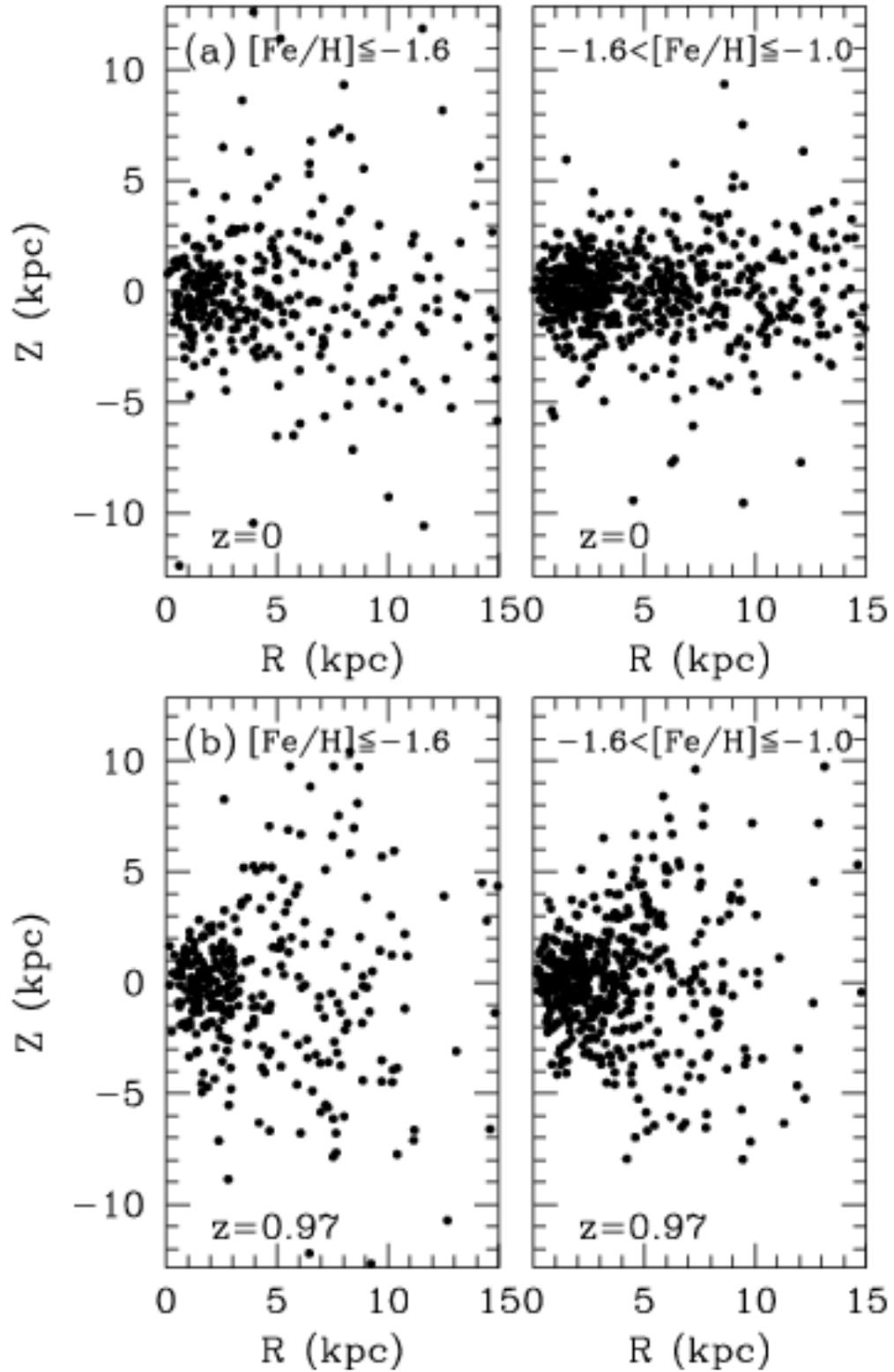}
\figcaption{
The same as Figure 11 but for the stars within $R = 15$ kpc and $|Z| = 15$ kpc.
\label{fig-12}}
\end{figure}

\begin{figure}
\epsscale{1.0}
\plotone{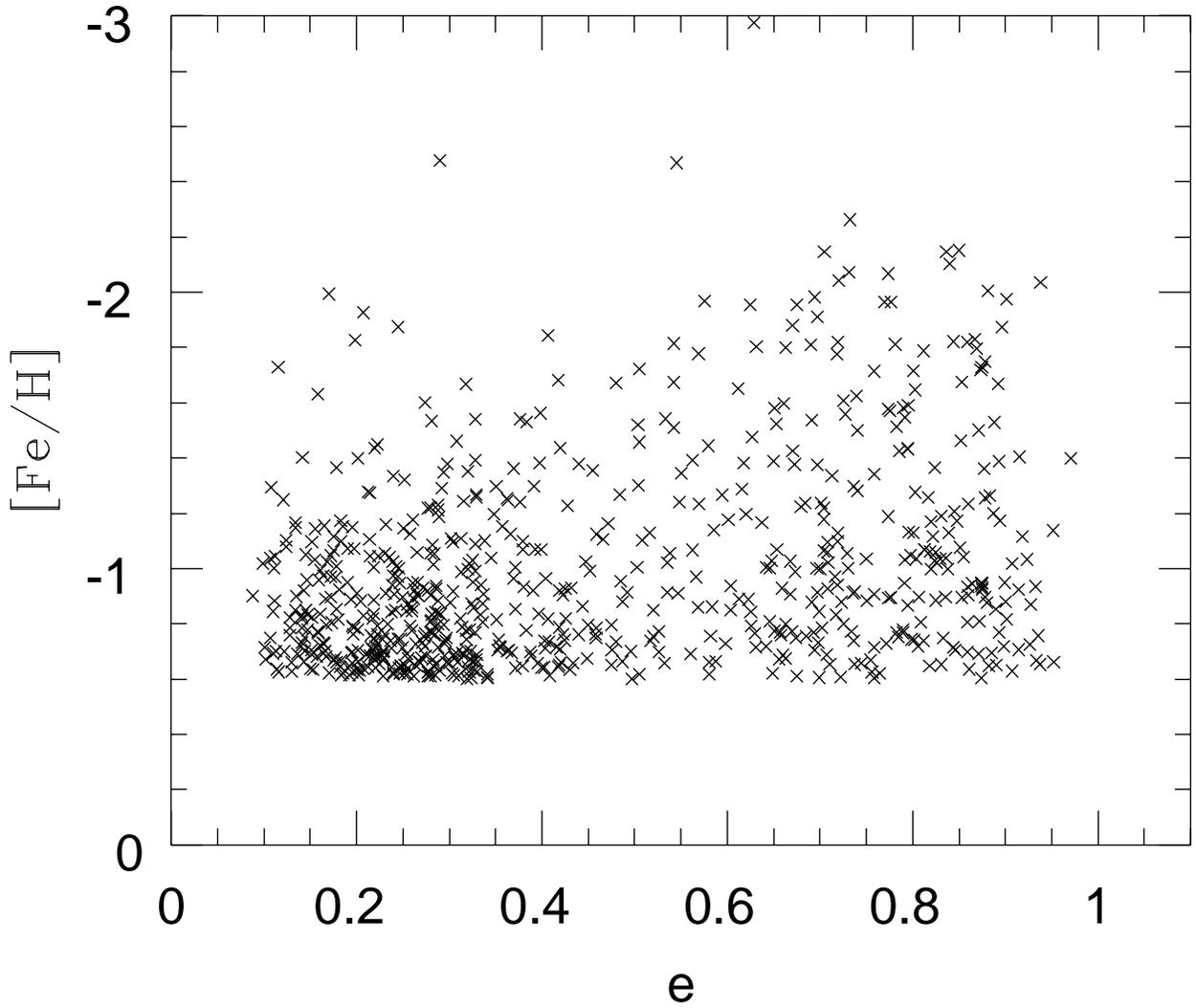}
\figcaption{
Relation between metal abundances ([Fe/H]) and orbital eccentricities ($e$)
for the simulated metal-poor stars with [Fe/H]$\le-0.6$.
Note the existence of low-$e$, low-[Fe/H] stars in the simulated halo.
\label{fig-13}}
\end{figure}

\begin{figure}
\epsscale{1.0}
\plotone{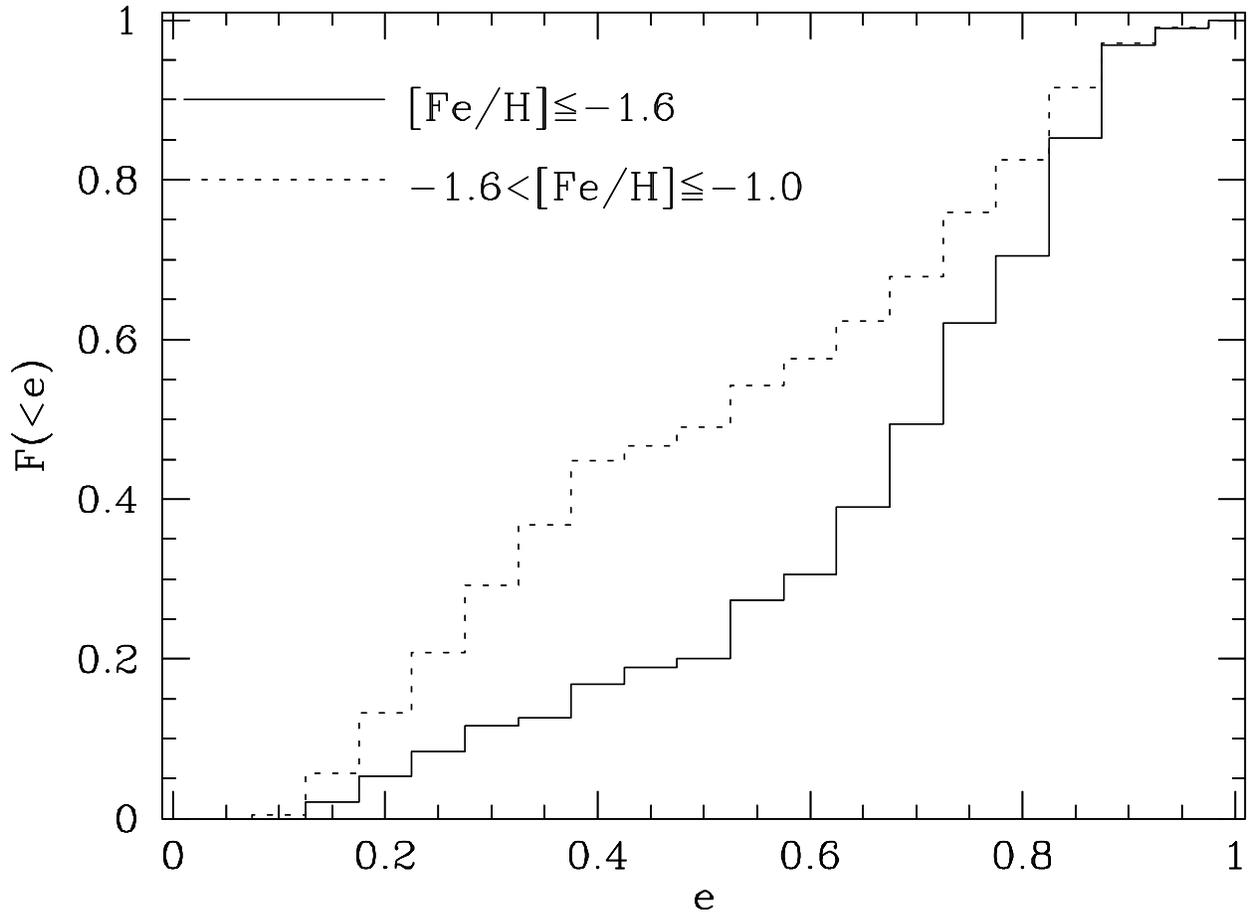}
\figcaption{
Cumulative $e$ distributions, $F(<e)$, in the two abundance
ranges of [Fe/H]$\le-1.6$ (solid line) 
and $-1.6<$[Fe/H]$\le-1.0$ (dotted one).
\label{fig-14}}
\end{figure}

\begin{figure}
\epsscale{1.0}
\plotone{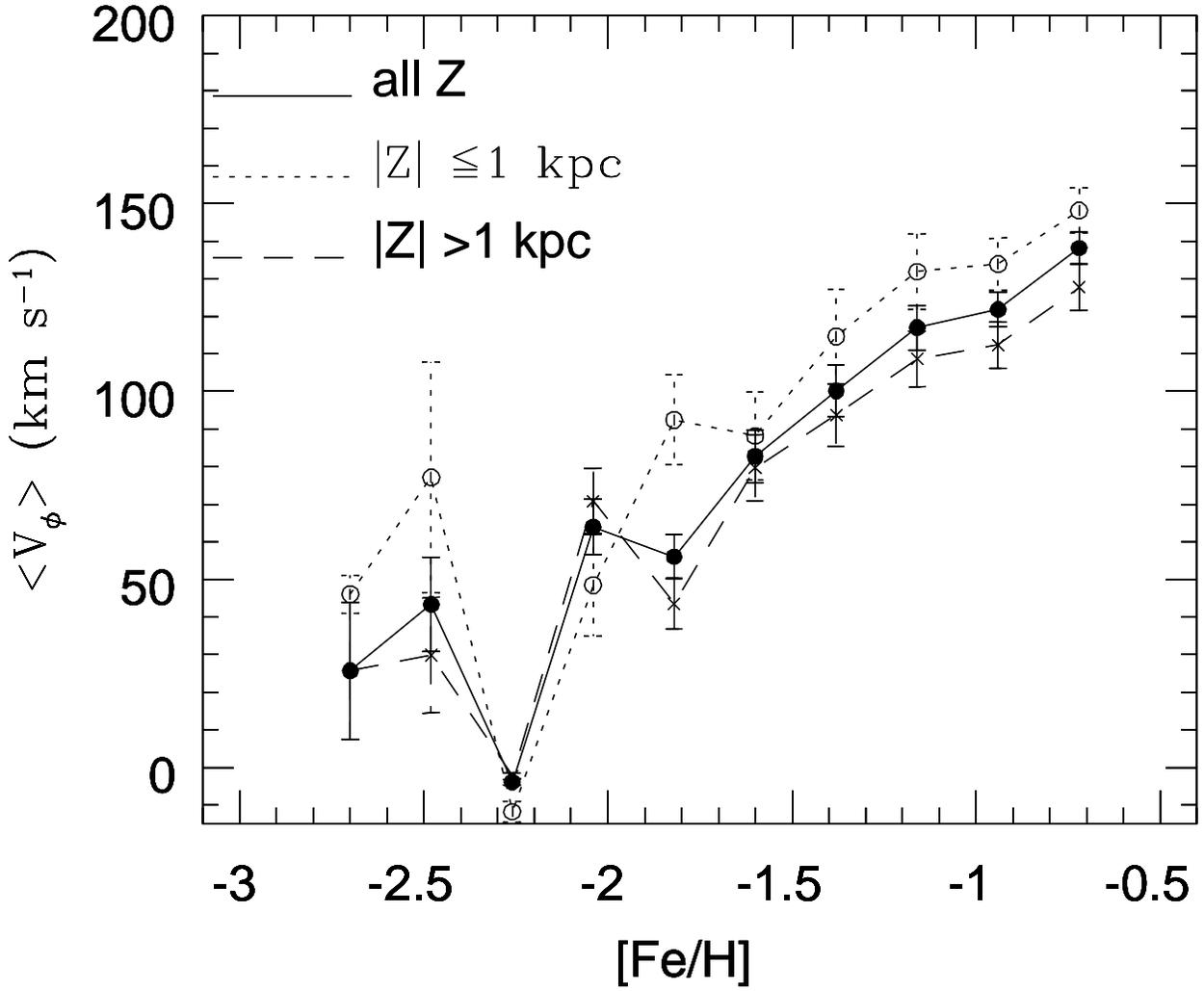}
\figcaption{
Mean rotational velocity $<V_\phi>$ vs. metal abundance [Fe/H], for
all stars (solid line), stars at $|Z|\le1$ kpc (dotted line),
and at $|Z|>1$ kpc (dashed line), where $Z$ is the distance from
the Galactic plane. Error  bars are plotted for each component.
\label{fig-15}}
\end{figure}

\begin{figure}
\epsscale{1.0}
\plotone{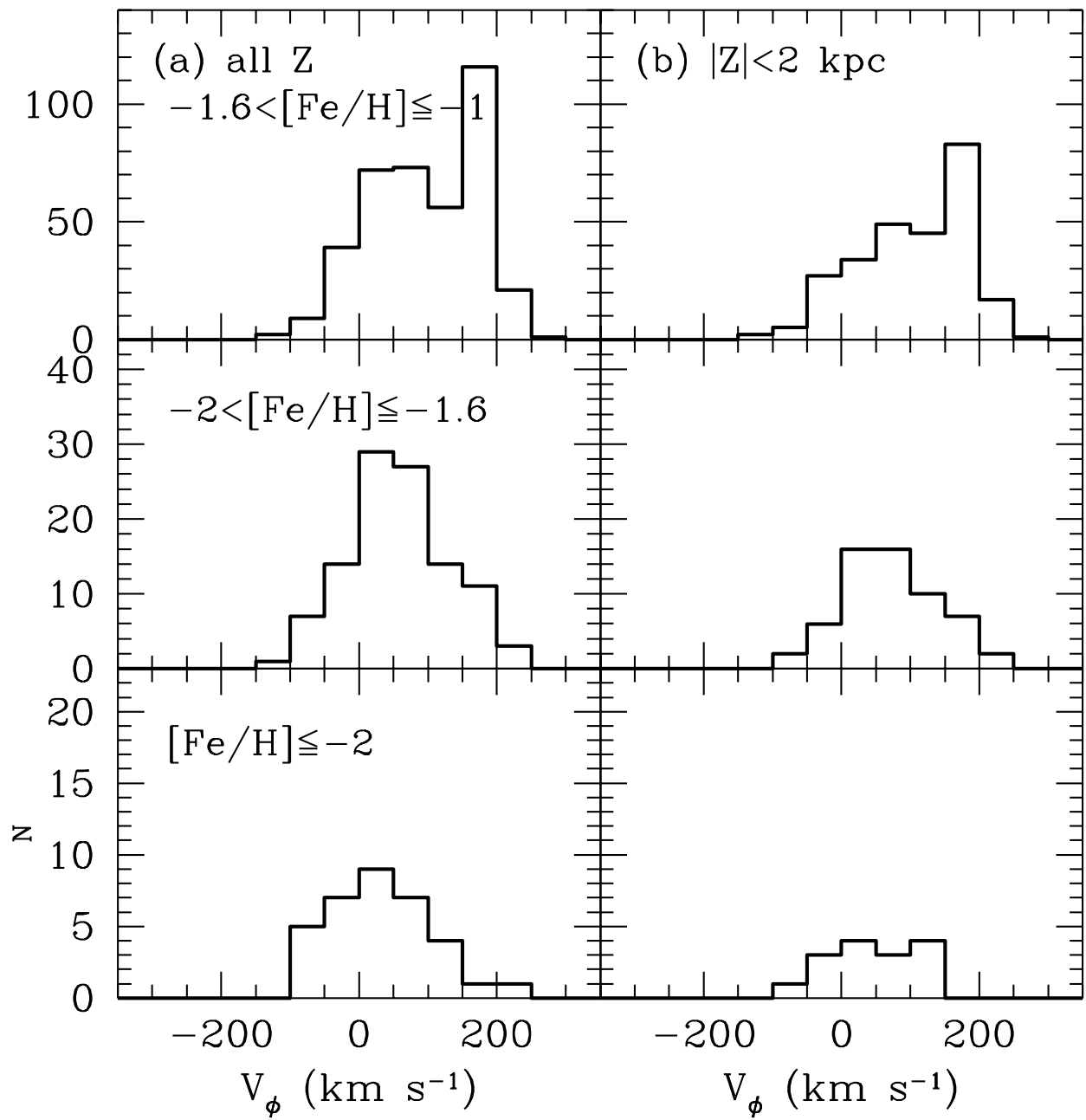}
\figcaption{
Distribution of $V_\phi$ in different metallicity ranges for the stars
at $R > 3$ kpc. Left and right panels denote stars at any $|Z|$
(left panels) and at $|Z|$ $\ge$ 2 kpc (right panels), respectively. 
\label{fig-16}}
\end{figure}

\begin{figure}
\epsscale{1.0}
\plotone{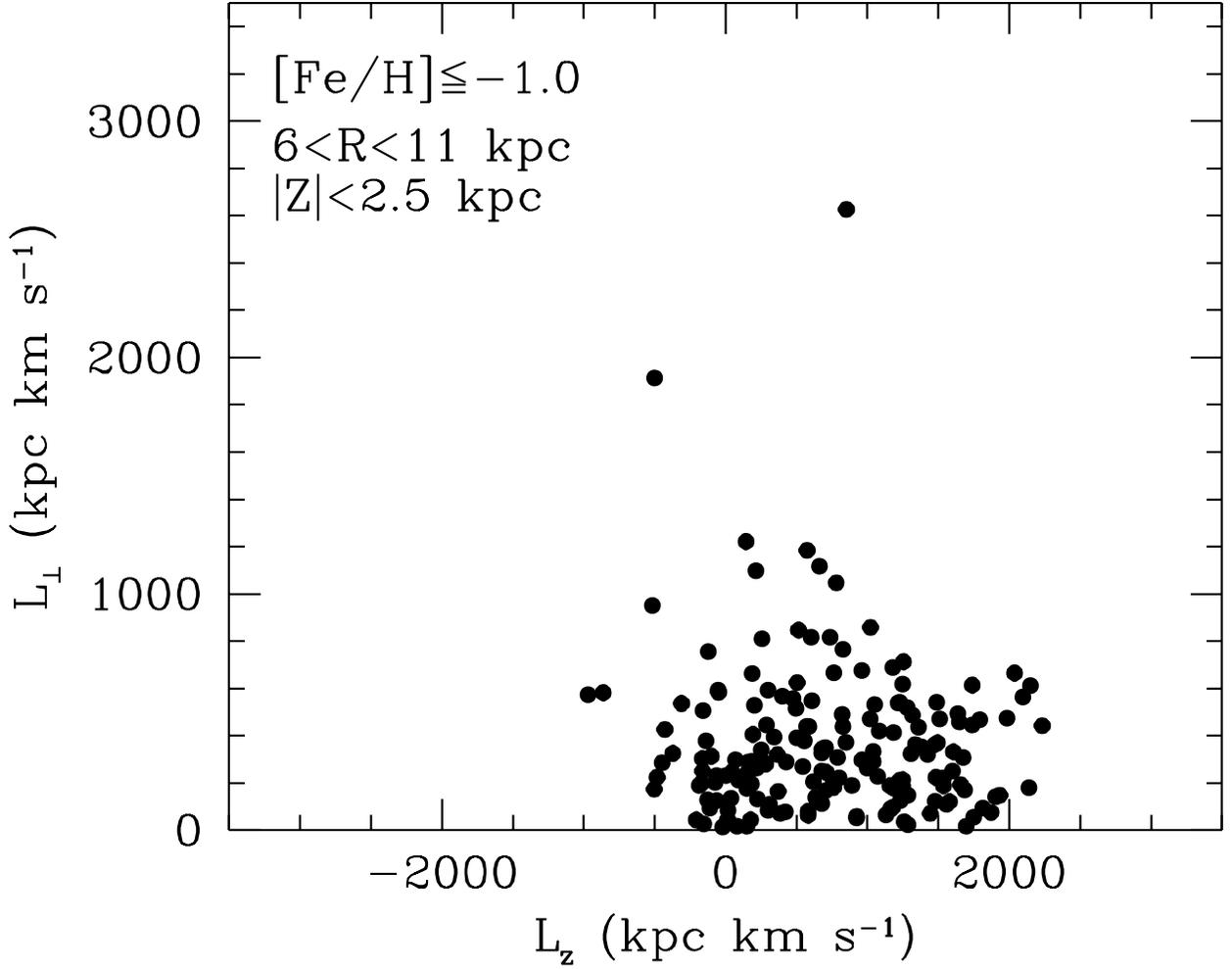}
\figcaption{
The present-day distribution of the simulated stars with [Fe/H] $\le$ $-1.0$
in angular momentum diagram $L_{\rm z}$ vs. $L_{\perp}=
{({L_{\rm x}}^2+{L_{\rm y}}^2)}^{1/2}$. We have selected the stars
at $6<R<11$ kpc and $|Z|<2.5$ kpc, to compare with the observational
result near the Sun.
\label{fig-17}}
\end{figure}

\begin{figure}
\epsscale{1.0}
\plotone{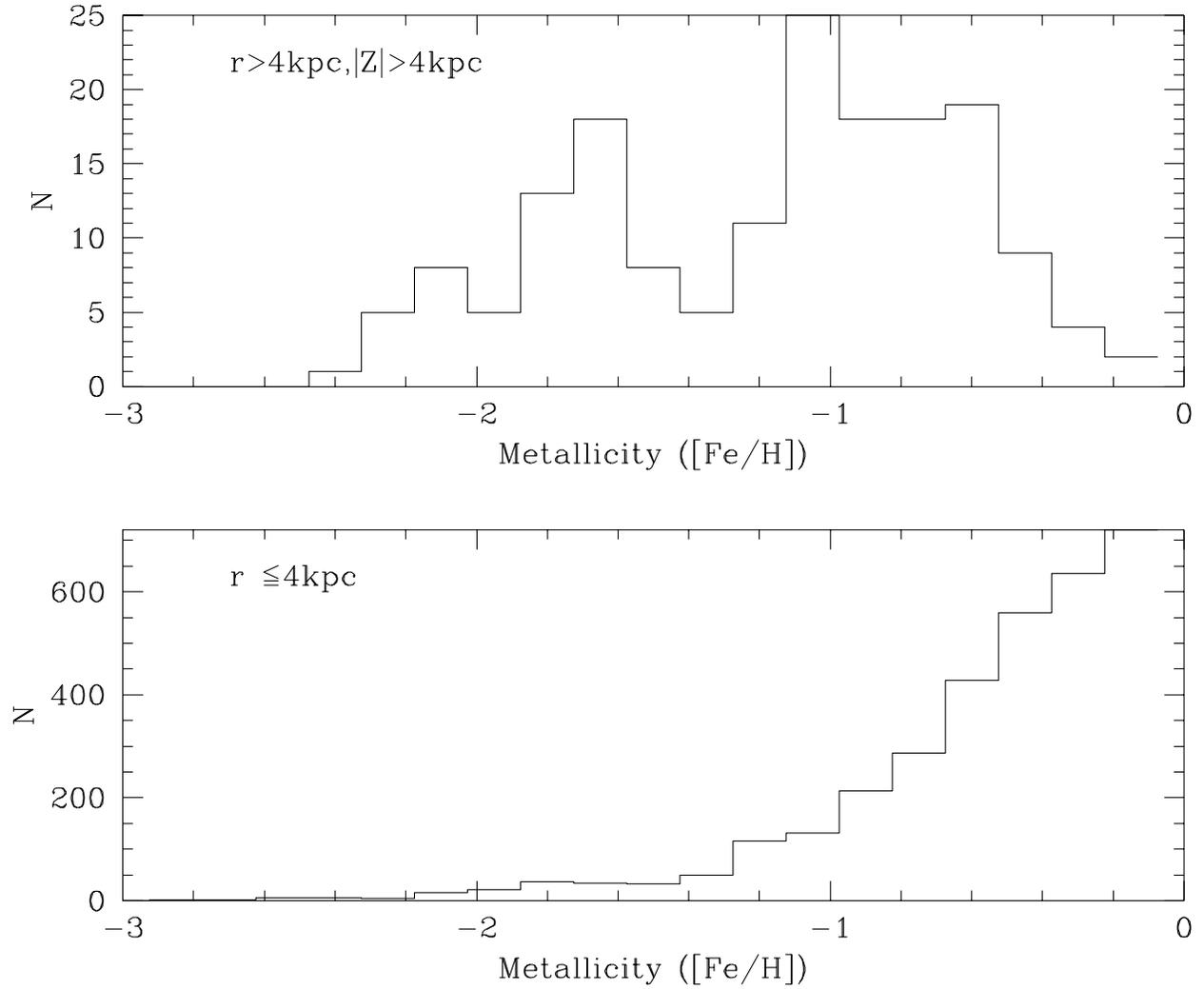}
\figcaption{
Metallicity distributions of the simulated stars at $r>4$ kpc and $|Z|>4$ kpc
(upper panel) and at $r\le 4$ kpc (lower panel), respectively. These panels
are to be compared with the observed metallicity distributions of the halo
and bulge, respectively.
\label{fig-18}}
\end{figure}

\clearpage

\begin{figure}
\epsscale{0.6}
\plotone{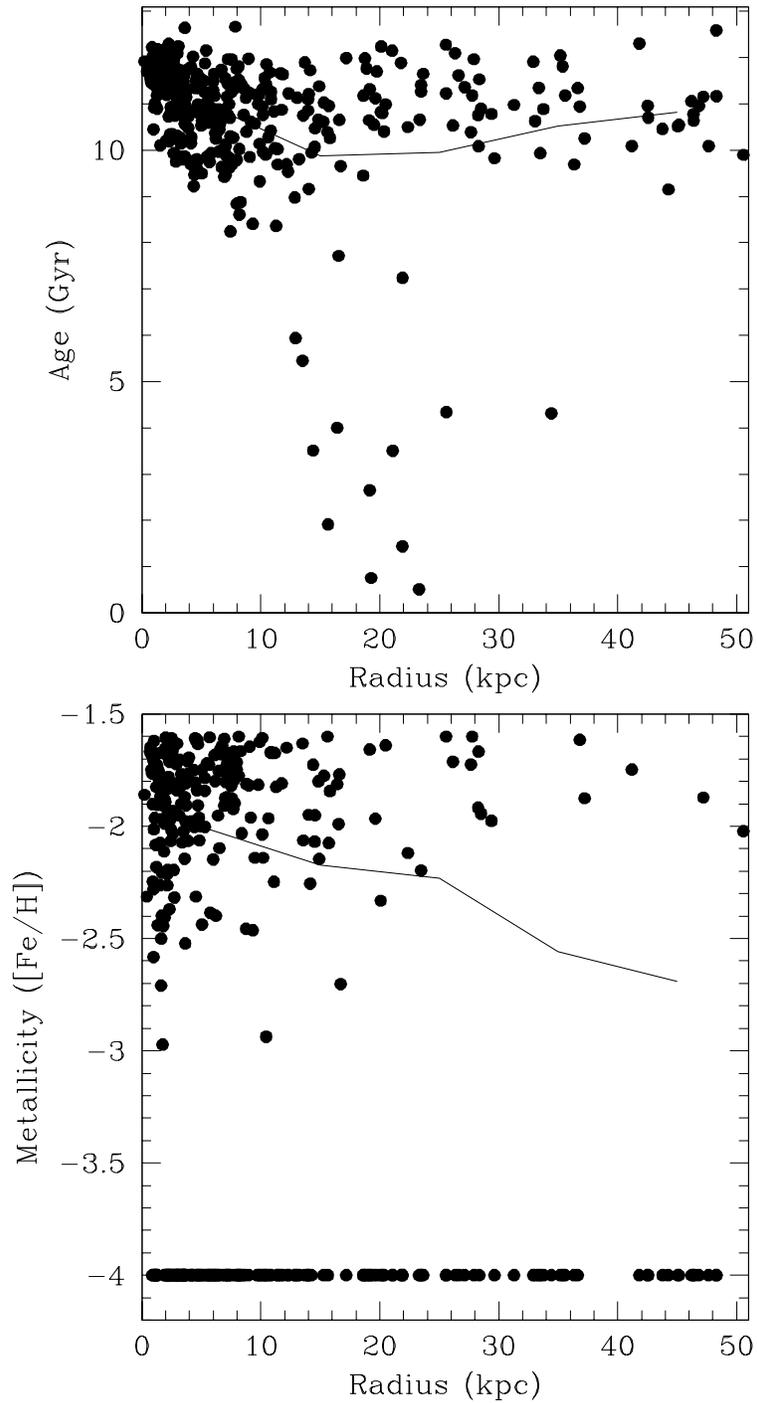}
\figcaption{
Radial distribution of stellar ages (upper panel) and metallicities
(lower panel) for the stars with [Fe/H] $\le$ $-1.6$.
Solid lines show local regression lines through the points.
For convenience, all of the stars with [Fe/H]$<-4$ are plotted
at [Fe/H] $=$ $-4$.
\label{fig-19}}
\end{figure}

\begin{figure}
\epsscale{1.0}
\plotone{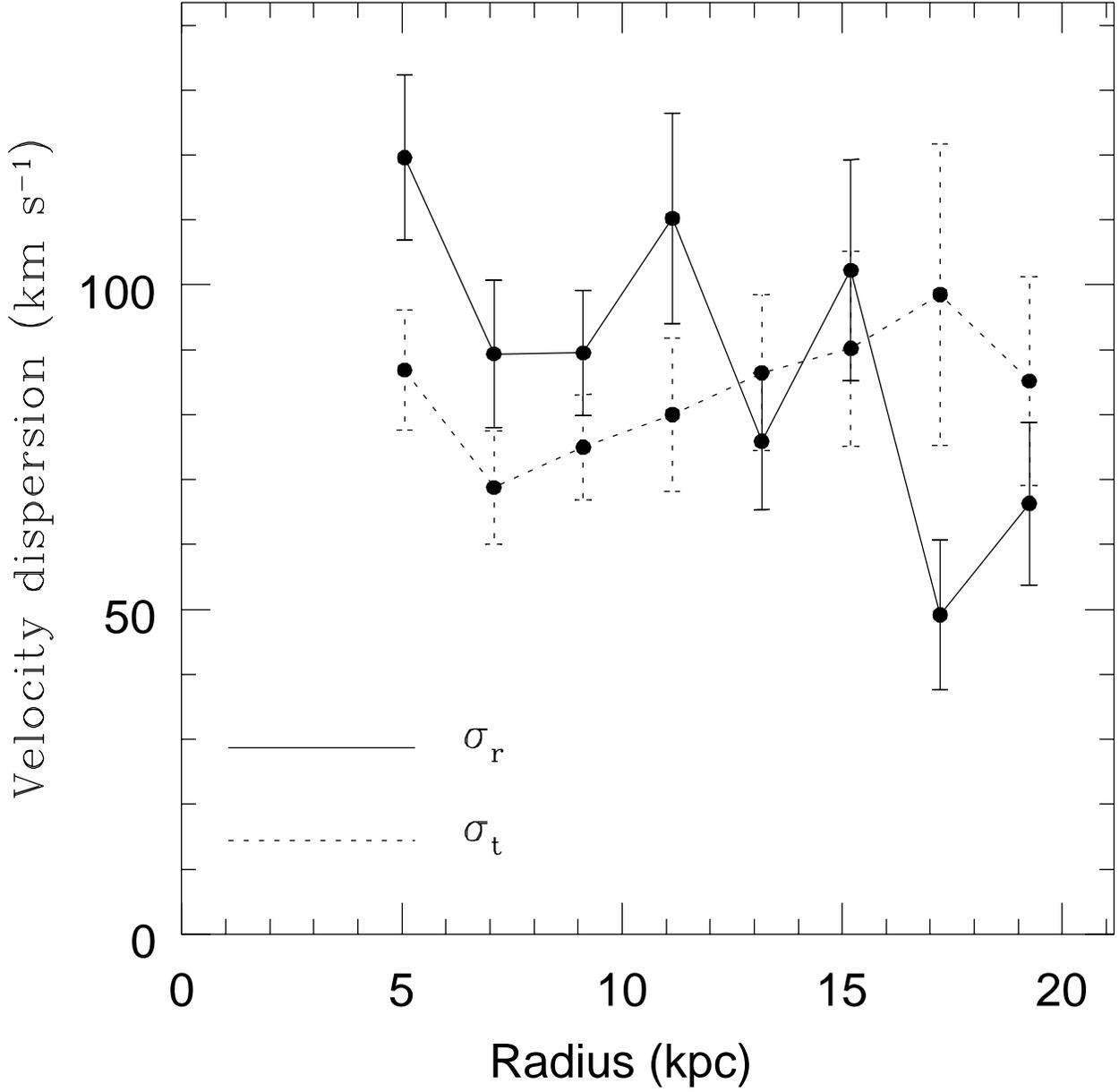}
\figcaption{
Radial dependence of radial velocity dispersion ${\sigma}_{\rm r}$ (solid line)
and tangential one  ${\sigma}_{\rm t}$  (dotted one) for the stars
with [Fe/H] $\le$ $-1.0$ at $|Z|>2$ kpc.
Error  bars are plotted for each component.
\label{fig-20}}
\end{figure}

\end{document}